\documentclass[a4paper,11pt]{article}
\usepackage{jcappub} % for details on the use of the package, please see the JINST-author-manual
\usepackage{graphicx}	% Including figure files
\usepackage{amsmath, physics}	% Advanced maths commands
\usepackage{hyperref, amsfonts, aas_macros}
\usepackage{soul}

\newcommand{\hi}{H\,{\textsc i} }

\newcommand{\blue}[1]{#1}

\title{\boldmath Seeing Wiggles without Seeing Wiggles: BAO Recovery in 21 cm Intensity Mapping with Deep Learning}

\author[a]{Kaifeng Yu}
\author[a,b]{and Xin Wang}
\affiliation[a]{School of Physics and Astronomy, Sun Yat-Sen University, Zhuhai, 519082, China}
\affiliation[b]{CSST Science Center for the Guangdong–Hong Kong–Macau Greater Bay Area, Sun Yat-Sen University, Zhuhai, 519082, China}

\emailAdd{yukf@mail.sysu.edu.cn}
\emailAdd{wangxin35@mail.sysu.edu.cn}

\abstract{
The $21\,\mathrm{cm}$ intensity mapping provides a promising probe of the large-scale structure. Astrophysical foregrounds, as the main source of contamination to the cosmological $21\,\mathrm{cm}$ signal, persist in a wedge-like region of Fourier space due to the inherent chromaticity in radio interferometric observations. The foreground avoidance strategy focuses on utilizing data from relatively clean regions with minimal foreground leakage, at the cost of losing large-scale information. 
Non-linear structure formation, however, couples Fourier modes across scales, leaving imprints of the missing large-scale modes in the remaining data. In this work, we employ a deep learning approach \blue{based on Convolutional Neural Networks (CNNs)} to test whether large-scale features of the $21\,\mathrm{cm}$ brightness temperature fields, particularly the baryon acoustic oscillations (BAO), can be recovered at the field level using only short-wavelength modes that are beyond the linear scales. To explicitly assess the dependence on the training cosmology, we train the network exclusively on de-wiggled simulations, providing a controlled test of whether the reconstruction arises from physical non-linear mode coupling rather than implicit encoding of BAO features.
In the ideal noise-free case, the amplitude and phase of the lost modes can be restored with high fidelity. With instrumental noise included, the reconstructed amplitude becomes biased, while the phase information remains robust. The trained network also exhibits reasonable robustness to variations in the underlying cosmological model. Together, these results suggest that mode restoration offers a complementary approach for extracting cosmological information from future $21\,\mathrm{cm}$ intensity mapping analyses.
}


\begin{document}
\maketitle
\flushbottom

\section{Introduction}
Hydrogen is the most abundant element in the Universe. Observations of the \blue{spatial distribution of the} redshifted $21\,\mathrm{cm}$ signal from atomic hydrogen \blue{enable the construction of three-dimensional tomographic maps, providing} a powerful probe and a unified view of cosmic structure formation and thermal evolution from the Dark Ages to low redshift \cite{2006PhR...433..181F, 2010ARA&A..48..127M, 2012RPPh...75h6901P, 2020PASP..132f2001L}.
In the era of post-reionization, most of the hydrogen is ionized, almost all of the $21\,\mathrm{cm}$ radiation comes from the self-shielded regions such as the galaxy, making it a promising tracer for the matter distribution. \hi intensity mapping (IM) is a technique for observing the integral $21\,\mathrm{cm}$ line emission over a wide area of the sky without resolving individual galaxies, enabling high survey efficiency and making it well suited for studying cosmological large-scale structure. One prominent application is the measurement of baryon acoustic oscillations (BAO) \cite{2008MNRAS.383.1195W, 2017MNRAS.466.2736V}, which has traditionally been carried out using galaxy surveys. In recent years, rapid and substantial advances in both observational capabilities and data analysis, evidenced by the successes of several experiments, have also demonstrated the potential of this technique \cite{2010Natur.466..463C, 2013ApJ...763L..20M, 2023MNRAS.518.6262C, 2023ApJ...947...16A, 2025MNRAS.537.3632M, 2025MNRAS.541..476M, 2025arXiv251119620C}.

The astrophysical foregrounds, including galactic synchrotron radiation, free-free emission, bright and unresolved extragalactic point sources, make the measurement of the cosmological $21\,\mathrm{cm}$ signal challenging. The foreground is about 4 to 6 orders of magnitude brighter than the $21\,\mathrm{cm}$ signal, such that even small residual contamination may overwhelm its detection.
The dominant foreground emission is expected to be spectrally smooth, which is often regarded as a promising property for separating it from the spectrally \blue{fluctuating} cosmological $21\,\mathrm{cm}$ signal. 
Although various foreground subtraction methods have been proposed and successfully validated using mock data (e.g. \cite{2015MNRAS.447..400A, 2018MNRAS.478.3640M, 2022MNRAS.509.2048S}), complex instrumental effects and systematic uncertainties, such as calibration errors and polarization leakage, make the practical implementation of these approaches nontrivial.
In practice, foreground subtraction can partially remove the underlying $21\,\mathrm{cm}$ signal, while residual foregrounds may persist in the subtracted data and introduce bias.
Specifically, the inherent chromatic responses of radio interferometers can induce mode mixing, causing foreground power to leak from intrinsically low-$k_{\parallel}$ modes into a broader region of Fourier space and resulting in the characteristic wedge-like foreground contamination \cite{2012ApJ...752..137M, 2013ApJ...770..156H,2014PhRvD..90b3018L}. This behavior motivates an alternative strategy, known as \textit{foreground avoidance}, which exploits the fact that foreground contamination is largely confined to a specific region in Fourier space. Rather than attempting to subtract the foregrounds, this strategy restricts power spectrum estimation to modes expected to be relatively foreground-free. Foreground avoidance has therefore been widely adopted, either alone or in combination with foreground subtraction in the analysis of $21\,\mathrm{cm}$ data from experiments such as HERA, MWA, and MeerKAT (e.g. \cite{2022ApJ...925..221A,2025MNRAS.541..476M,2025ApJ...989...57N}). The trade-off, however, is a loss of sensitivity in the recovered $21\,\mathrm{cm}$ power spectrum, particularly on the large scales relevant to BAO measurements.

Non-linear structure formation couples Fourier modes across scales, transferring information between scales and inducing correlations between modes.  Consequently, it is, in principle, feasible to recover large-scale information from small-scale modes of any tracer of the matter field. As mentioned above, this is particularly relevant for the measurement of $21\,\mathrm{cm}$ signal. Motivated by this, a number of studies have explored whether information outside the foreground wedge can be used to recover contaminated large-scale modes in the $21\,\mathrm{cm}$ fields. These efforts either perform reconstructions directly at the field level or aim to infer relevant astrophysical parameters. A variety of approaches have been explored, including the use of neural networks, the extension of a Lagrangian bias model, a formalism based on the effective field theory (EFT) framework, and other effective heuristic techniques \cite{2019JCAP...11..023M,2021MNRAS.504.4716G,2022MNRAS.509.3852P,2024MNRAS.529.3684K, 2025MLS&T...6a5039S,2025JCAP...04..082L, 2025ApJ...983..166L,2025JCAP...11..082C,2025arXiv250813268Q}.

In this paper, we constructed a deep learning model and applied it on simulation data to reconstruct the $21\,\mathrm{cm}$ brightness temperature fields from the  modes-removed $21\,\mathrm{cm}$ tomographic data. 
One of the main potential drawbacks of most machine learning based reconstruction methods is that they are inherently data-driven, and their performance may critically depend on the quality and coverage of the training set. It is often unclear whether the model has genuinely learned the underlying physical processes of non-linear mode coupling or whether it is primarily interpolating within the limited parameter space spanned by the training set. As a result, the degree to which the reconstructed signals depend on the assumed training cosmologies remains an open question.
Motivated by this concern, in this work we deliberately train our mode-recovery network using mock data of which the initial conditions are generated from a de-wiggled linear power spectrum. By completely removing the BAO-related Fourier modes from the training data, we design a controlled test to assess the extent to which the method is able to recover the BAO signal when applied to a realistic universe. \blue{When evaluated on fields that contain BAO features, the model successfully recovers the BAO wiggles, as seen in the power spectra of the recovered fields, suggesting that it captures aspects of the underlying mode coupling of the density fields.}

In Section \ref{sec:sim}, we detail the process of generating simulation data. In Section \ref{sec:nn}, we describe the neural network architecture and the training procedure applied in this work. In Section \ref{sec:result}, we present and discuss the results. We adopt the cosmological parameters from Planck 2018 \cite{2020A&A...641A...1P} as our fiducial model: the matter density $\Omega_{\rm m} = 0.309$, the baryon density $\Omega_{\rm b} = 0.049$, the Hubble constant $h = 0.6766$ and the amplitude of matter density fluctuations $\sigma_8 = 0.81$.

%%%%%%%%%%
\section{Simulation}
\label{sec:sim}

\subsection{Fast N-body Simulation}
\label{sec:nbody-sim}

Accurate predictions for the cosmic structure formation, particularly on the non-linear scales, need computationally expensive $N$-body simulations, making it impractical for generating a large amount of mock data.
The COmoving Lagrangian Acceleration (COLA) method uses a combination of second-order Lagrangian Perturbation Theory (2LPT) and particle mesh (PM) solver to enable fast approximate simulations. By using a reduced number of time steps to solve the $N$-body equations of motion, COLA captures the essential non-linear evolution while maintaining high fidelity on large-scale structures \cite{2013JCAP...06..036T, 2016MNRAS.459.2118K}. Although the COLA method lacks accuracy on small scales, it is sufficient for our purposes.

As mentioned previously, we aim to verify that the neural network is capable of capturing the mode coupling rather than merely memorizing specific patterns in the training data. To this end, BAO-related information is deliberately removed during training by de-wiggling the input power spectrum. 
The original linear matter power spectrum is obtained from the transfer function calculated using \texttt{CLASS} code \cite{2011JCAP...07..034B} wrapped in \texttt{nbodykit}\footnote{\url{https://github.com/bccp/nbodykit}} assuming the Planck 2018 cosmology, and the de-wiggled smooth power spectrum is then obtained by applying a Savitzky-Golay filter with a \blue{fourth}-order polynomial.

We employ \texttt{COLA-HALO}\footnote{\url{https://github.com/junkoda/cola_halo}} \cite{2016MNRAS.459.2118K} for fast $N$-body simulations to produce the dark matter density fields, all simulations were run with $512^3$ particles in a box of $(1024\, h^{-1} {\rm Mpc})^3$, each particle mass is about $6.9 \times 10^{11} \, h^{-1} M_{\odot}$. \blue{The initial conditions are generated at $z_{\rm init} = 9$ with 2LPT, and the dark matter fields are evolved to $z = 0$ with $10$ time steps uniformly spaced in scale factor $a$. The particle positions and velocities of our field snapshots at $z = 1 \,(a = 0.5)$ are obtained by interpolation between the adjacent steps.}
Then we apply the halo finder \texttt{RockStar}\footnote{\url{https://bitbucket.org/pbehroozi/rockstar-galaxies}} \cite{2013ApJ...762..109B} with linking length $0.2$ and minimum number of particles $20$ for halo finding.
We generated 200 realizations of COLA simulations at redshift $z = 1$ with the de-wiggled power spectrum using different initial condition random seeds for our network training and testing, and some simulation samples sharing the same initial condition seeds with the original linear power spectrum were also produced.

%%%%%%%%%%%%%%%%
\subsection{HI Mass and Brightness Temperature}
\label{sec:HI-Tb}

The minimal mass of the halos found in our simulation is $\sim 1.38 \times 10^{12} \,h^{-1} M_{\odot}$, this means that a massive amount of low mass halos hosting \hi are not identified. We use a conditional mass function model to populate the mock dark matter density fields with halos below the mass threshold (see also \cite{2016JCAP...03..001S, 2025JCAP...04..082L}), then assign the \hi mass to each halo following an empirical relation between \hi mass and halo mass via 
\begin{equation}
    M_{\rm \hi}(M; z) = M_0 \left( \frac{M}{M_\mathrm{min}(z)} \right)^{\alpha(z)} e^{-(M_\mathrm{min}(z) / M)^{0.35}},
\label{eq:halo-hi-mass}
\end{equation}
where $M_0$ determines the overall normalization, $M_{\rm min}(z)$ controls the cutoff mass around which the halos host a significant amount of neutral hydrogen. At $z = 1$, the best fit parameters are $M_0 = 1.5 \times 10^{10} \, h^{-1} M_{\odot}$, $M_\mathrm{min} = 6.0 \times 10^{11} \, h^{-1} M_{\odot}$, and $\alpha = 0.53$ \cite{2018ApJ...866..135V}.

The conditional mass function $n(m,\, z_1\, |\, M,\, V,\, z_0)$ gives the number density of halos with mass $m$ virialized at $z_1$ in a cell of volume $V$ containing mass $M$ at redshift $z_0$,  which can be modeled with \cite{2002PhR...372....1C}
\begin{equation}
    n(m,\, z_1\, |\, M,\, V,\, z_0) = \frac{\bar{\rho}}{m^2} \nu_{10} f(\nu_{10}) \frac{\dd \ln \nu_{10}}{\dd \ln m},
\end{equation}
where $\bar{\rho} = \Omega_\mathrm{m} \rho_\mathrm{crit}$ is the mean background matter density, and 
\begin{equation*}
    \nu_{10} \equiv \frac{[\delta_\mathrm{sc}(z_1) - \delta_{0}(\delta,\, z_0)]^2}{\sigma^2(m) - \sigma^2(M)},
\end{equation*}
in which $\delta_{\rm sc}(z) \approx 1.68647 / D(z)$ is the critical overdensity for a spherical collapse, $D(z)$ is the linear growth factor normalized to $1$ at $z = 0$. $\delta_{0}(\delta, z_0)$ denotes the initial density for a region with density $\delta = M/\bar{\rho}V - 1$ at $z_0$, which is extrapolated with linear theory. An approximation to relation between $\delta_0$ and $\delta$ for the spherical collapse model can be obtained by \citep{1996MNRAS.282..347M, 2002MNRAS.329...61S} 
\begin{equation*}
\begin{aligned}
    \delta_0(\delta,\,z)
    &= \frac{\delta_{\mathrm{sc}}(z)}{1.68647} \times \biggl[1.68647 - \frac{1.35}{(1+\delta)^{2/3}} -\frac{1.12431}{\left(1+\delta\right)^{1/2}}+\frac{0.78785}{\left(1+\delta\right)^{0.58661}}\biggr],
\end{aligned}
\end{equation*}
and $\sigma(m)$ and $\sigma(M)$ are the root mean square of the initial density fluctuation field in spheres containing mass $m$ and $M$. After converting to the Lagrangian radius $R$ for the halo mass $M$ using the relation $R = \left(\frac{3 M}{4 \pi \bar{\rho}_{m, 0}} \right)^{1/3}$, the corresponding $\sigma_m(R)$ or $\sigma_M(R)$ can be obtained via
\begin{equation*}
    \sigma^2(R, z) = \int \frac{\dd k}{k} \frac{k^3 P(k, z)}{2 \pi^2} |W(k R)|^2,
\end{equation*}
which means that the variance in the smoothed density field with a smoothing window $W$ of scale $R$. We take a top-hat window $W(k R) = [3 / (k R)^3] [\sin (k R) - k R \cos (k R)]$ here. And an analytic fit to $\nu f(\nu)$ is given by
\begin{equation*}
    \nu f(\nu) = A \left(1 + \frac{1}{(a \nu)^{p}}\right) \left( \frac{a \nu}{2 \pi} \right)^{1/2} e^{-a \nu / 2},
\end{equation*}
where $A = 0.3222$, $a = 0.75$, $p = 0.3$ \citep{2002MNRAS.329...61S}.

\begin{figure}[ht!]
    \centering
    \includegraphics[width=0.8\linewidth]{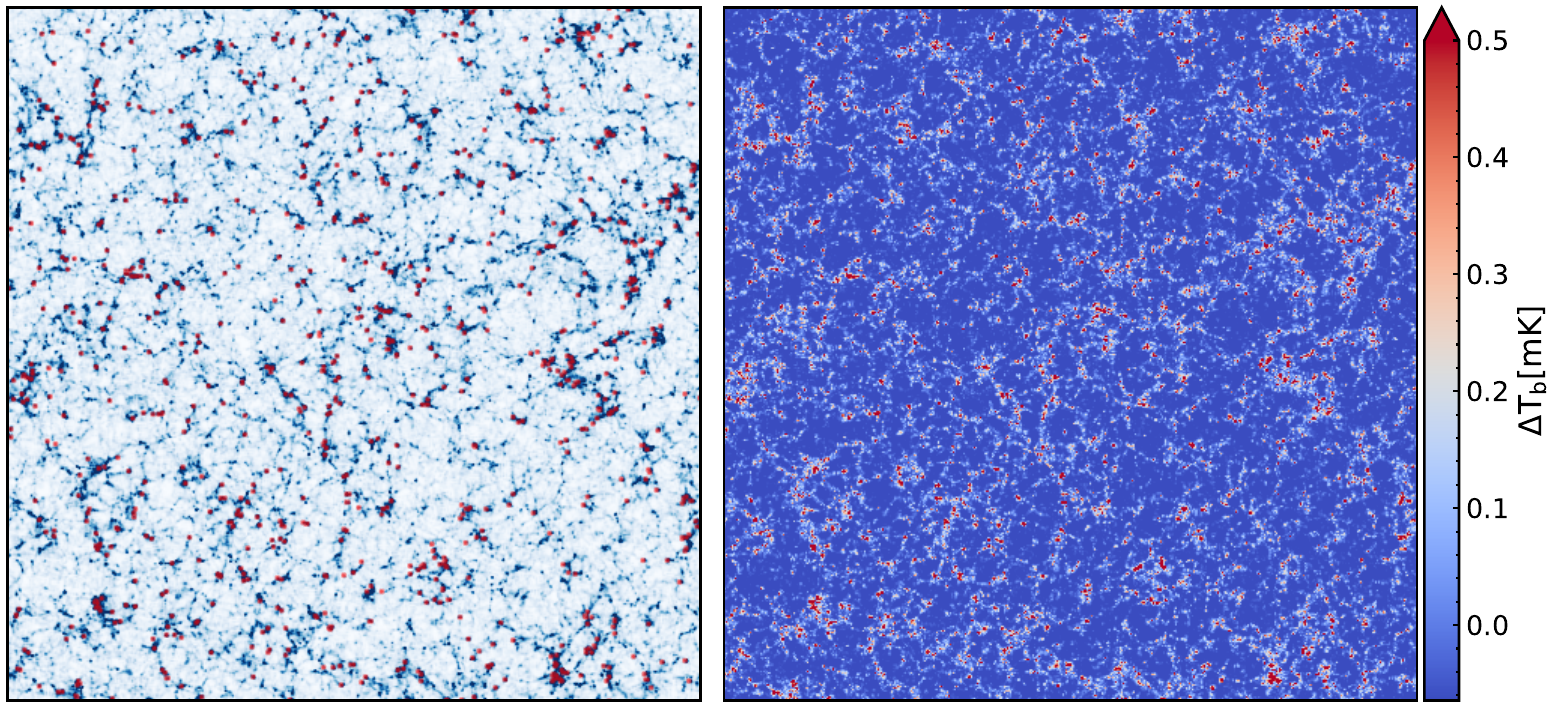}
    \caption{A slice of the dark matter distribution (\emph{left}) and the corresponding fluctuation of $21\,\mathrm{cm}$ brightness temperature (\emph{right}), taken in a plane transverse to the line-of-sight direction. The \emph{red} dots in the left panel indicate the position of halos.}
    \label{fig:dm-Tb}
\end{figure}

The number of halos $N_i$ within a mass bin $[m_i,\,m_i + \dd m_i] \in [m_{\rm min},\, m_{\rm max}]$ in a cell of mass $M$ is then sampled from a Poisson distribution
\begin{equation*}
    N_i \sim \mathrm{Pois}(n(m_i\, |\, M) V_{\rm cell}),
\end{equation*}
where $m_{\rm min}$ and $m_{\rm max}$ specify the minimum and maximum halo mass for the sub-grid halos sampling. We take $m_{\rm min}$ as the hard cutoff mass $M_{\rm hard}(z=1)= 6.9 \times 10^{9} \, h^{-1} M_{\odot}$ defined in \cite{2018ApJ...866..135V} and $m_{\rm max}$ as the minimum halo mass identified by halo finder. 

We apply the Triangular Shaped Cloud (TSC) mass assignment scheme to paint the dark matter density fields and halo catalogs onto the grid before calculating the \hi brightness temperature fields. To account for the effect of redshift space distortion (RSD), the position of a halo in redshift space is defined as
\begin{equation*}
    \vb*{s} = \vb*{r} + \frac{v_{\rm los}}{a H(a)} \hat{\vb*{n}},
\end{equation*}
where $\vb*{r}$ is the real position of the halo, $v_{\rm los}$ is the peculiar velocity along the line-of-sight $\hat{\vb*{n}}$, and $H(a)$ is the Hubble parameter at scale factor $a$. The effect of peculiar velocities of individual sub-grid halos is not treated separately. Instead, RSD effect has been applied on dark matter particles when computing the sub-grid halo population within each cell.

After obtaining the number of sampled sub-grid halos within each mass bin, combined with the halo catalogs mesh, the total \hi mass in each cell is calculated with Equation \eqref{eq:halo-hi-mass}. The brightness temperature of the cell is then obtained by \cite{2015ApJ...803...21B, 2022MNRAS.510.3495W}
\begin{equation}
    T_{b}({\theta}, z)=\frac{3 h A_{12} c^3}{32 \pi m_{\mathrm{h}} k_{\mathrm{b}} \nu_{21}^2} \frac{(1 + z)^2}{H(z)} \frac{M_{\mathrm{\hi}}({\theta}, z)}{V_{\rm cell}},
\end{equation}
where $h$ is the Planck constant, $m_\mathrm{h}$ is the mass of hydrogen atom, $k_\mathrm{b}$ is the Boltzmann constant, $c$ is the speed of light, $\nu_{\rm 21}$ is the rest frequency of \hi emission line, $A_{12} \approx 2.869 \times 10^{-15} \, \mathrm{s}^{-1}$ is the Einstein coefficient for spontaneous emission from the $21\,\mathrm{cm}$ hyperfine transition, and $V_{\rm cell}$ denotes the comoving volume of the cell. The final $21\,\mathrm{cm}$ brightness temperature fields are produced by subtracting the mean value $\Delta T_b(\theta,\,z) = T_{b}(\theta,\, z) - \bar{T_b}(z)$.
For illustration, Figure~\ref{fig:dm-Tb} shows a realization of a slice of dark matter density field and the position of halos found by halo finder, as well as the corresponding $21\,\mathrm{cm}$ brightness temperature fluctuation field.

%%%%%
\subsection{Foreground Wedge}
\label{sec:fg-wedge}

The foreground emission in the cosmological $21\,\mathrm{cm}$ signal observations is expected to be spectrally smooth and therefore primarily contaminates the low-$k_{\parallel}$ modes in Fourier space. However, the frequency-dependent response of the interferometer leads to mode-mixing, \blue{which redistributes foreground power from low $k_{\parallel}$ to higher $k_{\parallel}$, with an extent that increases with $k_{\perp}$.} As a result, the foreground-contaminated modes occupy a characteristic wedge-like region in Fourier space, which can be bounded by a mathematical relation (e.g. \cite{2010ApJ...724..526D, 2012ApJ...752..137M, 2014arXiv1408.4695C, 2014PhRvD..90b3018L})
\begin{equation}
    k_\parallel \leq |\boldsymbol{k}_\perp| \sin\theta \frac{E(z)}{1+z}\int_0^z\frac{\mathrm{d}z^{\prime}}{E(z^{\prime})} + b.
\label{eq:fg-wedge}
\end{equation}
Here, $E(z) = \sqrt{\Omega_m (1 + z)^3 + \Omega_\Lambda}$, $k_\parallel$ and $\boldsymbol{k}_\perp$ denote the Fourier wavenumber of modes parallel and perpendicular to the line-of-sight, respectively. The parameter $b$ is introduced to characterize the width of the intrinsic foreground contamination and the instrumental bandwidth in low $k_{\parallel}$, $\theta$ represents the angular size of the field of view, we take it to be the horizon limit $\pi/2$ \blue{for the pessimistic scenario}. In this work, we adopt $b = 0.1$, and the resulting slope of the wedge boundary is approximately $0.68$ \blue{at $z = 1$}. We do not consider the case of a foreground-contaminated region larger than the conventional one described above \cite{2025A&A...693A.276M}. In addition to applying the analytic foreground-wedge removal, we further truncate modes with $k < 0.3 \,h{\rm Mpc}^{-1}$ to completely exclude the region of linear scales and most of the visible BAO features.

%%%%%
\subsection{Instrumental Noise}
\label{sec:noise}

In practical observations, instrumental systematics are unavoidable, and accurate modeling of these effects are crucial for the detection of the cosmological $21\,\mathrm{cm}$ signal. The sources of instrumental systematics in $21\,\mathrm{cm}$ observations include calibration errors, imperfect modeling of the primary beams, polarization leakage, and instrumental noise, etc. In addition, data loss caused by radio frequency interference (RFI), solar contamination during observation, and other observational artifacts further degrade the signal-to-noise ratio of the measurements.
In this work, we consider instrumental thermal noise as the only instrumental effect and neglect other sources of systematics. We assume an idealized, perfectly calibrated system with no data loss. Other effects that may impact imaging performance, such as primary beam corrections and errors associated with wide field imaging and mosaicing, are also neglected in this study.

\begin{figure}
    \centering
    \includegraphics[width=0.4\linewidth]{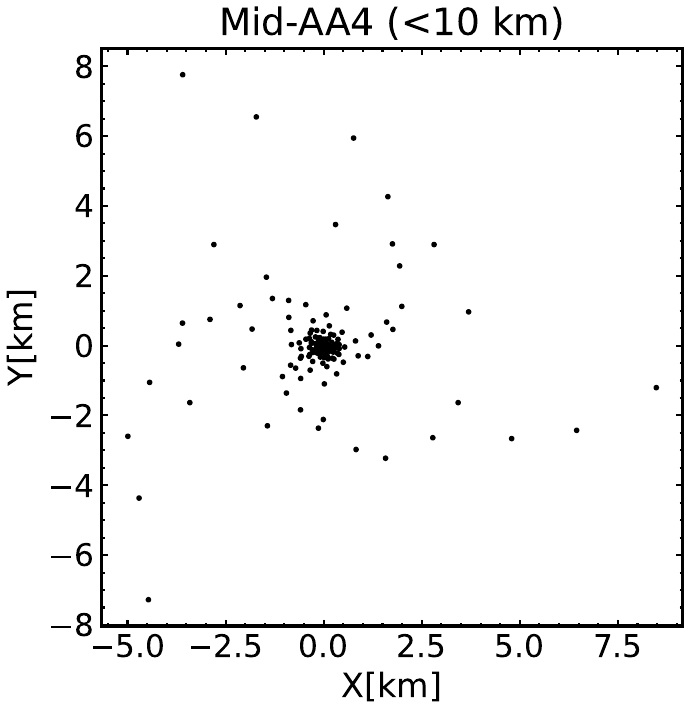}
    \hspace{0.2cm}
    \includegraphics[width=0.5\linewidth]{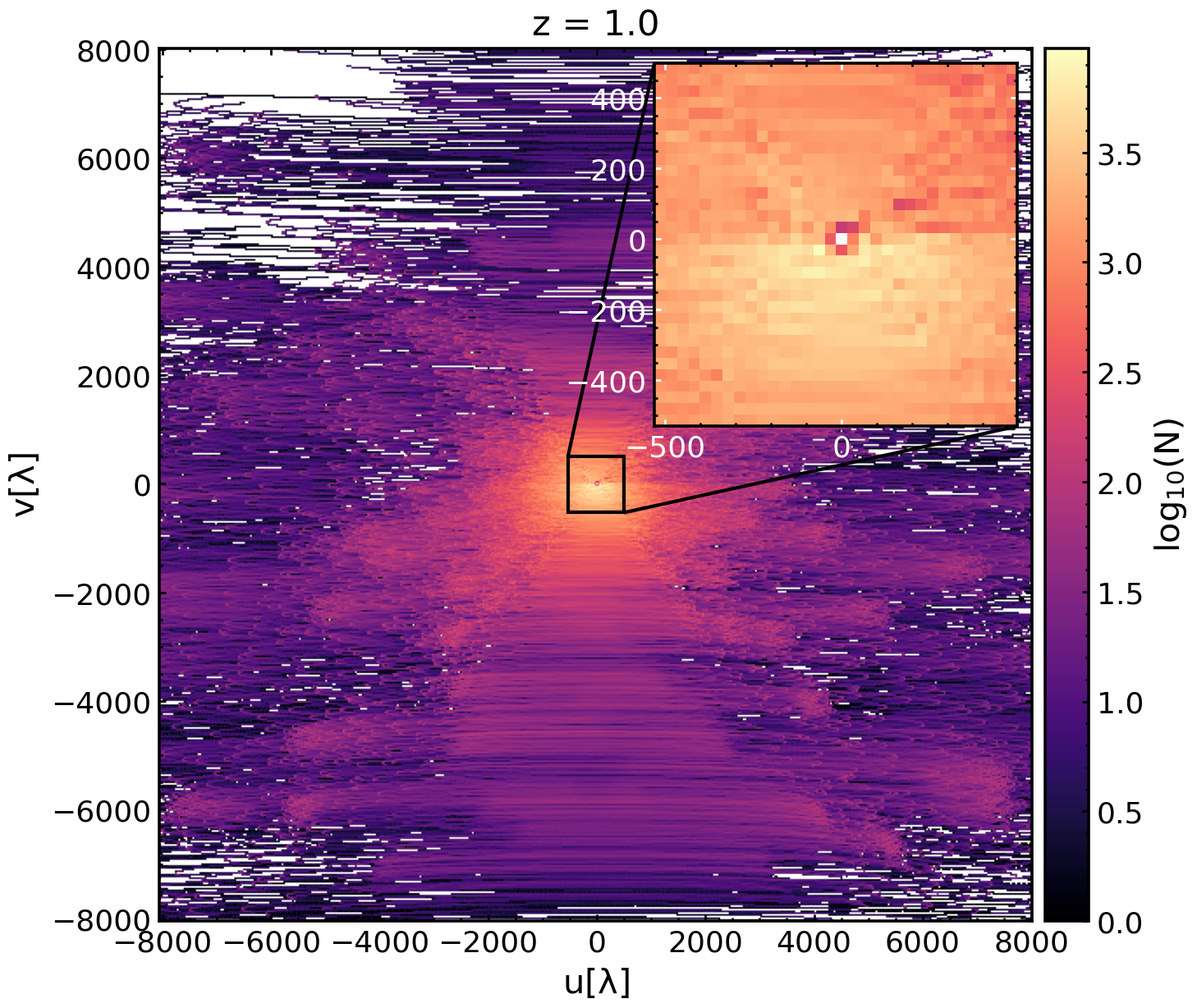}
    \caption{\emph{Left}: The configuration of 164 SKA-Mid~\texttt{AA4} antennas located within $10\,$km of the center of array. \emph{Right}: A portion of the baseline distribution in the \textit{uv} plane for the SKA-Mid~\texttt{AA4} configuration, assuming an $8\,$h tracking observation of the COSMOS field at $z=1.0$. Owing to the limited spatial resolution of our simulations, only a truncated region of the \textit{uv} plane, shown as the zoomed-in part area, is retained for the generation of thermal noise.}
    \label{fig:ska-mid-aa4}
\end{figure}

To demonstrate our result, we consider a representative cosmological survey with the upcoming Square Kilometre Array (SKA) (e.g. \cite{Maartens:2015mra, 2020PASA...37....7S}). For BAO-focused observations, the currently proposed SKA cosmology surveys primarily adopt the single-dish mode. However, this mode provides limited access to non-linear scales, which are essential for our analysis. 
We therefore take the SKA-Mid~\texttt{AA4} configuration as the observing instrument to generate mock thermal noise realizations for interferometric \hi intensity mapping surveys. The geographic location of the telescope and the array layout of SKA-Mid~\texttt{AA4} are produced using the \texttt{ska-ost-array-config}\footnote{\url{https://gitlab.com/ska-telescope/ost/ska-ost-array-config}} package. The left panel of Figure~\ref{fig:ska-mid-aa4} shows the distribution of antennas located within $10\,\mathrm{km}$ from the center of the array. 

The measurement in interferometric observations is called \textit{visibility}, which represents the \blue{cross-correlation} of the output voltages from a pair of antennas. \blue{It is defined as}
\begin{equation*}
    V_{pq} = \int I(\hat{\vb*{n}}) A_{pq}(\hat{\vb*{n}}) e^{-2\pi i \frac{\vb*{b}_{pq} \cdot \hat{\vb*{n}}}{\lambda} }  \dd^2 \hat{\vb*{n}},
\end{equation*}
\blue{where $I(\hat{\vb*{n}})$ is the sky intensity distribution, $A_{pq}(\hat{\vb*{n}}) = A_{p}(\hat{\vb*{n}}) A_{q}^{*}(\hat{\vb*{n}})$ denotes the beam response. The unit direction vector $\hat{\vb*{n}}$ gives the sky position, $\vb*{b}_{pq}$ is the spatial separation between antenna $p$ and antenna $q$, $\lambda$ is the observing wavelength. In the flat-sky approximation,}
\begin{equation*}
    V_{pq} \approx \int I(l,\, m) A_{pq}(l,\, m) e^{-2\pi i (ul + vm)} \dd l \dd m,
\end{equation*}
\blue{where the components $(u,\, v)$ of the baseline vector $\vb*{b}_{pq} / \lambda$ define the $uv$ plane, $l$ and $m$ are the projection of $\hat{\vb*{n}}$ onto the $u$- and $v$-axes, respectively.}

The distribution of antenna separations within the array affects the observational performance and the resulting signal-to-noise ratio. It also determines the accessible range of \blue{$k_{\perp} = 2\pi \sqrt{u^2 + v^2} / D_c$} in the power spectrum measurement, \blue{where $D_c$ is the transverse comoving distance.}
Assuming that the receiver thermal noise is Gaussian, the RMS of noise in the real or imaginary component of the Stocks-I visibilities is given by
\begin{equation}
    \sigma_{\rm S} = \frac{k_{\rm b}  T_{\rm sys}}{A_e \sqrt{\delta \nu \tau}} \,[{\rm jy}],
\end{equation}
where $A_e = \eta \pi D^2 / 4$ is the effective collecting area of a single dish. We adopt an aperture efficiency of $\eta = 0.81$ and a dish diameter of $D = 15.0\,{\rm m}$ here. At $z = 1$, the resolution of the simulation box along the line-of-sight corresponds to a frequency channel width of $\delta \nu \approx 426.5\,$kHz, and the integration time per visibility is assumed to be $\tau = 30\,{\rm s}$. The system temperature is given by $T_{\rm sys} = T_{\rm rx} + T_{\rm CMB} + T_{\rm spill} + T_{\rm sky} \approx 26\,{\rm K}$, where the receiver temperature is modeled as $T_{\rm rx} = 15\,{\rm K} + 30\,{\rm K} \left( {\nu}[{\rm GHz}] - 0.75\right)^2$, the CMB temperature is $T_{\rm CMB} = 2.73\,{\rm K}$, and $T_{\rm spill} \approx 3\,{\rm K}$ accounts for the contribution from spillover. The sky temperature, dominated by the Galactic emission, is modeled as $T_{\rm sky} \approx 25\,{\rm K} \left( {408} \,/\,{\nu[{\rm MHz}]} \right)^{2.75}$~\cite{2020PASA...37....7S}.

The RMS of the weighted noise visibility in cell \blue{$\vb*{u}_i = (u_i,\, v_i)$} is given by
\begin{align*}
    \sigma_{V_{\rm N}'} (\vb*{u}_i) = \sqrt{N(\vb*{u}_i)} \sigma_{\rm S}(\vb*{u}_i),
\end{align*}
where $N(\vb*{u}_i)$ denotes the number of visibilities in the cell, and we assume the visibility weight $W = 1$ for all samples, corresponding to the natural weighting in the interferometric imaging terminology. After drawing the random noise complex weighted visibilities $V_{\rm N}' (\vb*{u}_i)$ from distribution $\mathcal{N} (0,\ \sigma^2_{V_{\rm N}'} (\vb*{u}_i))$, the noise data in real space fields are then obtained by Fourier transforming the weighted visibilities
\begin{equation}
    T_{N}(\hat{\vb*{x}}) = \mathrm{IFFT}\left(\frac{V_{\rm N}' (\vb*{u})}{\sum_{i}N(\vb*{u}_i)}\right),
\end{equation}
where $\mathrm{IFFT}$ indicates the \blue{inverse} Fast Fourier transform operator, $\sum_{i}N(\vb*{u}_i)$ represents the total number of visibility samples within the specified region of \textit{uv} coverage, serving as a normalization factor. The simulated \textit{uv} coverage of an $8\,$h tracking observation of the COSMOS field is illustrated in the right panel of Figure~\ref{fig:ska-mid-aa4}. We have truncated the full coverage to concentrate on the denser region. The zoomed-in panel highlights the area that is consistent with the box resolution of our $N$-body \blue{simulations}. Since the \textit{uv} coverage is nearly complete and approximately uniform within the region of interest, we neglect the potential impact of side-lobes of the point spread function (PSF). Accordingly, we do not convolve the $21\,\mathrm{cm}$ temperature fields with the corresponding PSF here. The impact of PSF and other previously mentioned observational effects will be investigated in future work using end-to-end simulations.
In addition, the noise level in the final productions from real radio interferometric imaging would differ as a result of various data processing steps, the treatments adopted in this work are therefore simplified. We expect that these simplifications do not affect the conclusions of our proof-of-concept study.

The field of view of each observation is approximated by $\lambda / D \approx 0.028\,{\rm rad}$, which corresponds to approximately $32$ pixels in our simulation box. For the observation of a single field, the total observation time used to generate the random noise is $400$ hours. The noise cube is produced by constructing $16 \times 16$ independent sets of noise data, each with dimensions $32 \times 32 \times 512$, where $512$ corresponds to the frequency axis. These sets together form a complete noise realization of the full datacube.

%%%%%%%%%%
\section{Neural Network and Training}
\label{sec:nn}

\begin{figure}[htbp]
    \centering
    \includegraphics[width=0.95\linewidth]{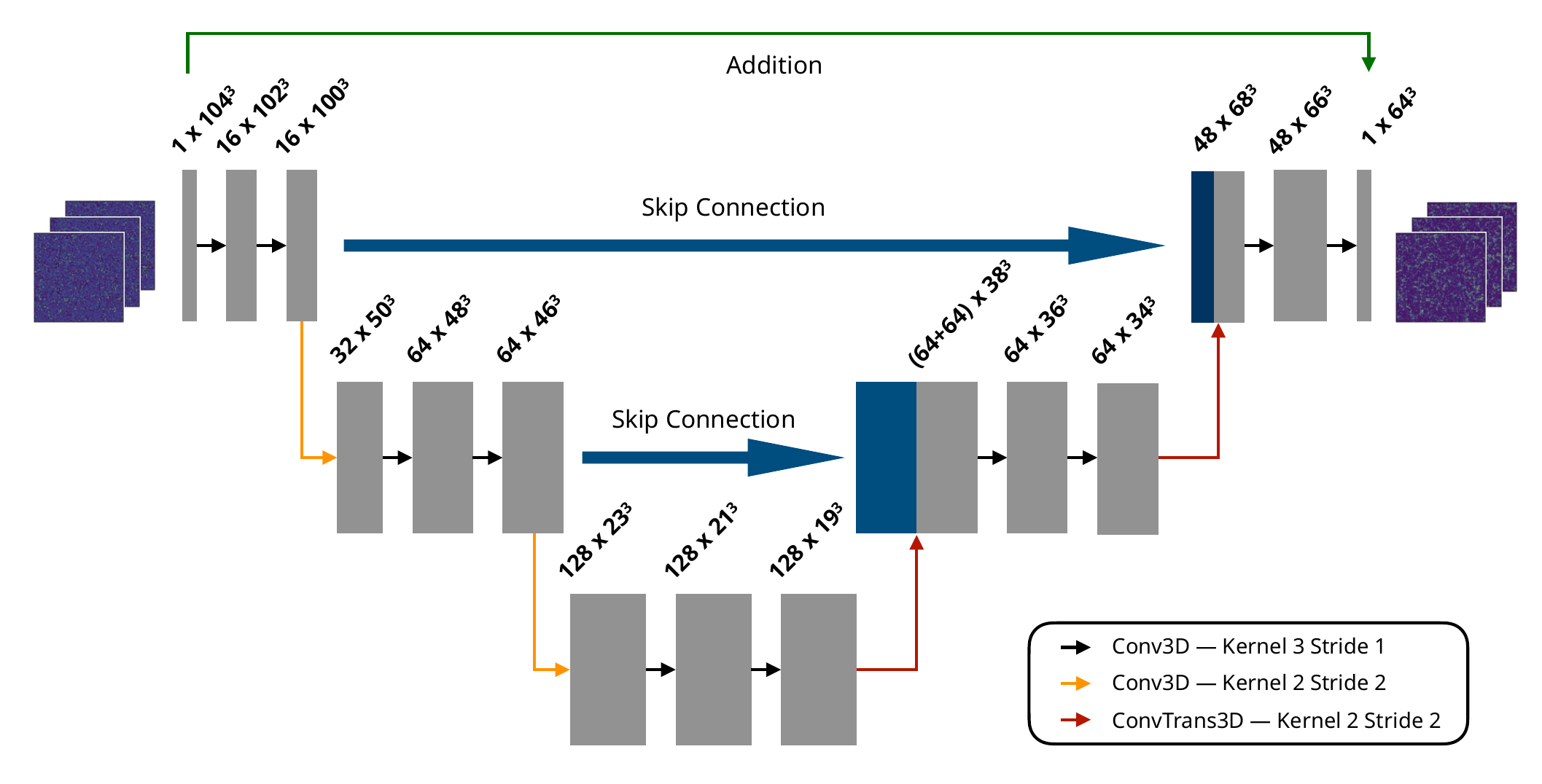}
    \caption{Diagram of the neural network architecture. The network follows an encoder-decoder structure in which the fields after removing specific modes are processed through successive 3D convolution layers (\emph{black} arrows) and down-sampling layers (\emph{orange} arrows). The decoder employs 3D transposed convolutions layers (\emph{red} arrows) to perform up-sampling operations, and incorporates skip connections (\emph{blue} arrows) through concatenation. A global residual connection (\emph{green} arrow) performs element-wise addition between the input and network output. Before being passed to the network, each datacube is padded from a size of $64^3$ to $104^3$. The network output has a dimension of $64^3$ and therefore requires no additional cropping.}
    \label{fig:nn-architecture}
\end{figure}

Convolutional Neural Networks (CNNs) are widely used in image processing and analysis, with the U-Net architecture demonstrating particular efficacy in extracting multi-scale features. The network model implemented in this work adopts a 3D U-Net-style architecture, which is similar to the framework presented in \cite{2019PNAS..11613825H} and illustrated in Figure~\ref{fig:nn-architecture}.

The network is designed to learn the residual between the input data (namely the modes removed fields) and the target data (the noise-free fields without modes removal). The architecture follows an encoder-decoder structure connected via concatenation layers. The encoding path comprises two convolution blocks, each consisting of two successive 3D convolution layer with a $3^3$ kernel and a stride of $1$. The down-sampling layer is then implemented using a 3D convolution operator with a $2^3$ kernel and a stride of $2$. Conversely, the decoding path employs 3D transposed convolution operators \blue{with the same kernel size and stride} to perform up-sampling and progressively recover the spatial resolution. 
With the exception of the first and last convolution layers, all convolution operators in the network are followed by group normalization and a Leaky ReLU activation function with a negative slope of $0.01$.

It is worth noting that all down-sampling (up-sampling) layers employ convolution (transposed convolution) rather than the commonly used max pooling or average pooling (interpolation-based up-sampling algorithms). 
\blue{Without padding, convolution reduces the size of the output. To ensure that the network output matches the target dimensions while preserving translation equivariance, we apply padding to the input data prior to the convolutional layers. In particular, following \cite{2023ApJ...952..145J}, we adopt periodic boundary padding applied only at the input stage. 
By contrast, commonly used padding schemes in machine learning, such as constant or reflective padding, would break the continuity of the field across the boundaries of the simulation box~\citep{2019PNAS..11613825H}. }

During the training phase, we only use the de-wiggled simulations as mentioned before. The $200$ simulated datacubes are divided into $180$ training sets and $20$ testing sets, separate models are trained for the noise-free and noisy cases. Due to the limitations of GPU memory, each full $512^3$ cube is split into $8^3$ subcubes of size $64^3$. Padding is then applied to each subcube\footnote{Only the subcubes located at the boundaries of the box are padded using periodic boundary conditions, whereas padding for interior subcubes is achieved by slicing larger subcubes from the original box.}, resulting in $104^3$ subcubes as network input. During training, a batch of 10 subcubes is passed to the network at each iteration. The loss function is defined as the mean squared error (MSE) between the network output datacubes of size $64^3$ and the corresponding true $21\,\mathrm{cm}$ brightness temperature fields without removing the specific modes. An Adam optimizer with a fixed learning rate $10^{-4}$ is applied. After obtaining the network output for each subcube, the results are then reassembled according to their indices in the original box to recover the full datacube.

%%%%%%%%%%
\section{Results and Discussion}
\label{sec:result}

\begin{figure}
    \centering
    \includegraphics[width=0.95\linewidth]{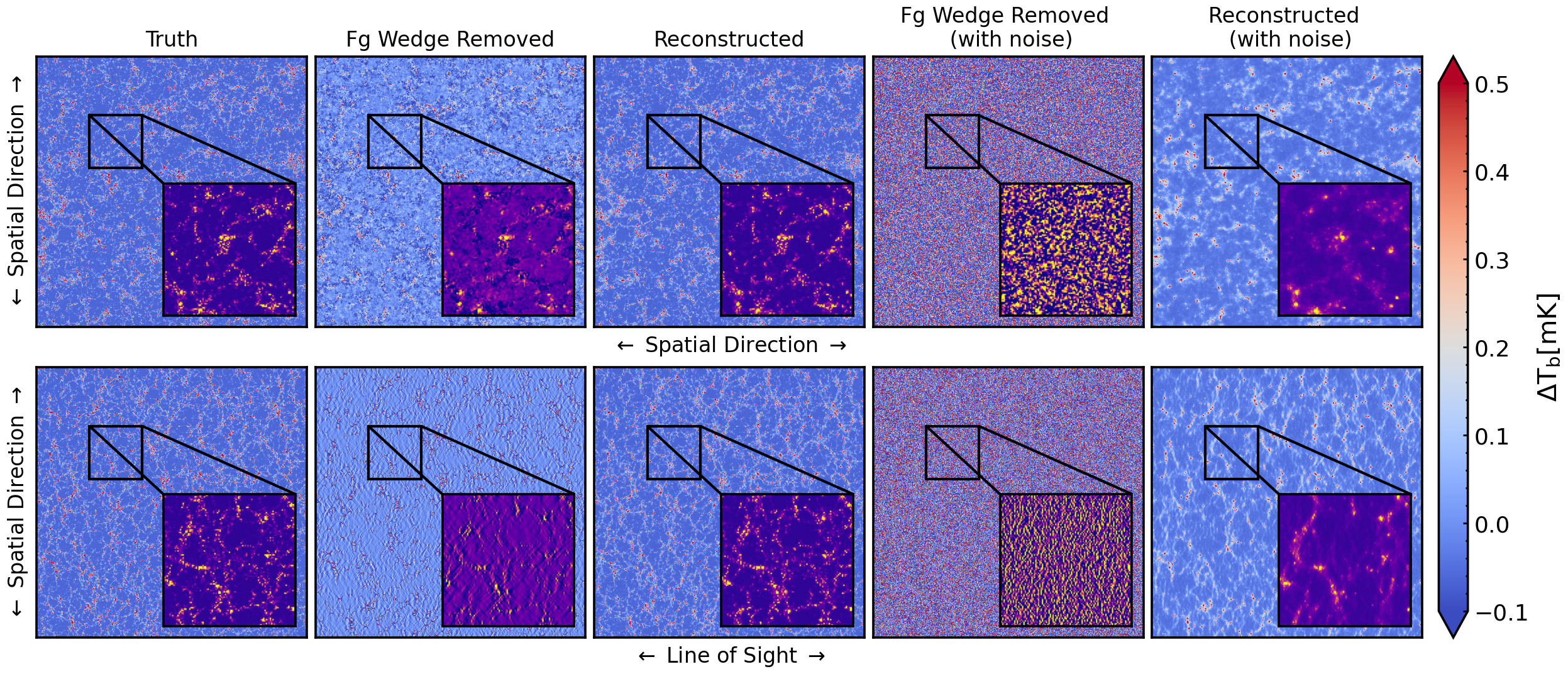}
    \caption{Slices of the $21\,\mathrm{cm}$ brightness temperature fields with volume of $(1024\, h^{-1} {\rm Mpc})^3$. The \emph{top} row shows a slice along the direction of line-of-sight, the \emph{bottom} row shows a slice along a spatial direction. The \emph{first} column shows the $21\,\mathrm{cm}$ field without removing modes in the foreground wedge and linear scales. The \emph{second} and \emph{third} columns show the field after removing the specific modes and the corresponding neural network reconstruction for the noise-free case. The \emph{last two} columns display the modes removed field and reconstructed field for the case including noise. In each panel, a selected region is zoomed in to more clearly illustrate the reconstruction performance.}
    \label{fig:nn-recon}
\end{figure}

As an illustration, Figure~\ref{fig:nn-recon} shows the underlying $21\,\mathrm{cm}$ field without specific modes removal, as well as the fields used as the network input with modes removal. After removing these modes, the structure of the field exhibits significant distortion both transverse to and along the line-of-sight. The corresponding reconstructed fields are also shown.
As can be seen, for the ideal noise-free case the field can be reconstructed with high fidelity. In the presence of observational noise, even when the signal is not visually discernible at low signal-to-noise ratio, the trained network remains capable of denoising and reconstructing the input field. Nevertheless, it is evident that the reconstructed field loses small-scale information and fine structures due to the presence of noise, resulting in an output that resembles a low-pass filtered version of the true field.
In the following, we evaluate our reconstruction model using standard cosmological statistics, including the power spectrum, transfer function, and cross-correlation coefficient, rather than image-based metrics such as pixel-wise MSE or Structural Similarity Index (SSIM). Our analysis focuses on the recovery of BAO features that are intentionally excluded from the training data.

%%%%%
\subsection{Power Spectrum}

The power spectrum of the brightness temperature fluctuation field $P(\vb*{k})$ is defined as
\begin{equation*}
    (2\pi)^3\delta^{D}(\vb*{k} - \vb*{k}')P(\vb*{k}) = \left< \Delta T_{\rm b}(\vb*{k}) \Delta T_{\rm b}^*(\vb*{k}')\right>,
\end{equation*}
where $\delta^{D}$ is the Dirac delta function, $\Delta T_b(\vb*{k})$ is the comoving spatial Fourier transform of field $\Delta T_{\rm b}(\hat{\vb*{x}})$, and $\left< \cdot \right>$ denotes an ensemble average. To quantify the reconstruction performance in terms of amplitude and phase, we employ two metrics. One is the transfer function, defined as 
\begin{equation}
    T(\vb*{k}) = \sqrt{\frac{P_{\rm recon}(\vb*{k})}{P_{\rm in}(\vb*{k})}},
\end{equation}
which measures the recovery of Fourier-mode amplitudes, and the other is cross-correlation coefficient between the input and reconstructed fields
\begin{equation}
    r_{\rm in \times recon}(\vb*{k}) = \frac{P_{\rm in \times recon}(\vb*{k})}{\sqrt{P_{\rm in}(\vb*{k}) P_{\rm recon}(\vb*{k})}},
\end{equation}
which quantifies the phase alignment between the two fields. Here, $P_{\rm in \times recon}(k)$ denotes the cross power spectrum between the input and reconstructed fields, while $P_{\rm in}(k)$ and $P_{\rm recon}(k)$ represent the corresponding auto power spectrum, respectively.

\begin{figure}
    \centering
    \includegraphics[width=0.45\linewidth]{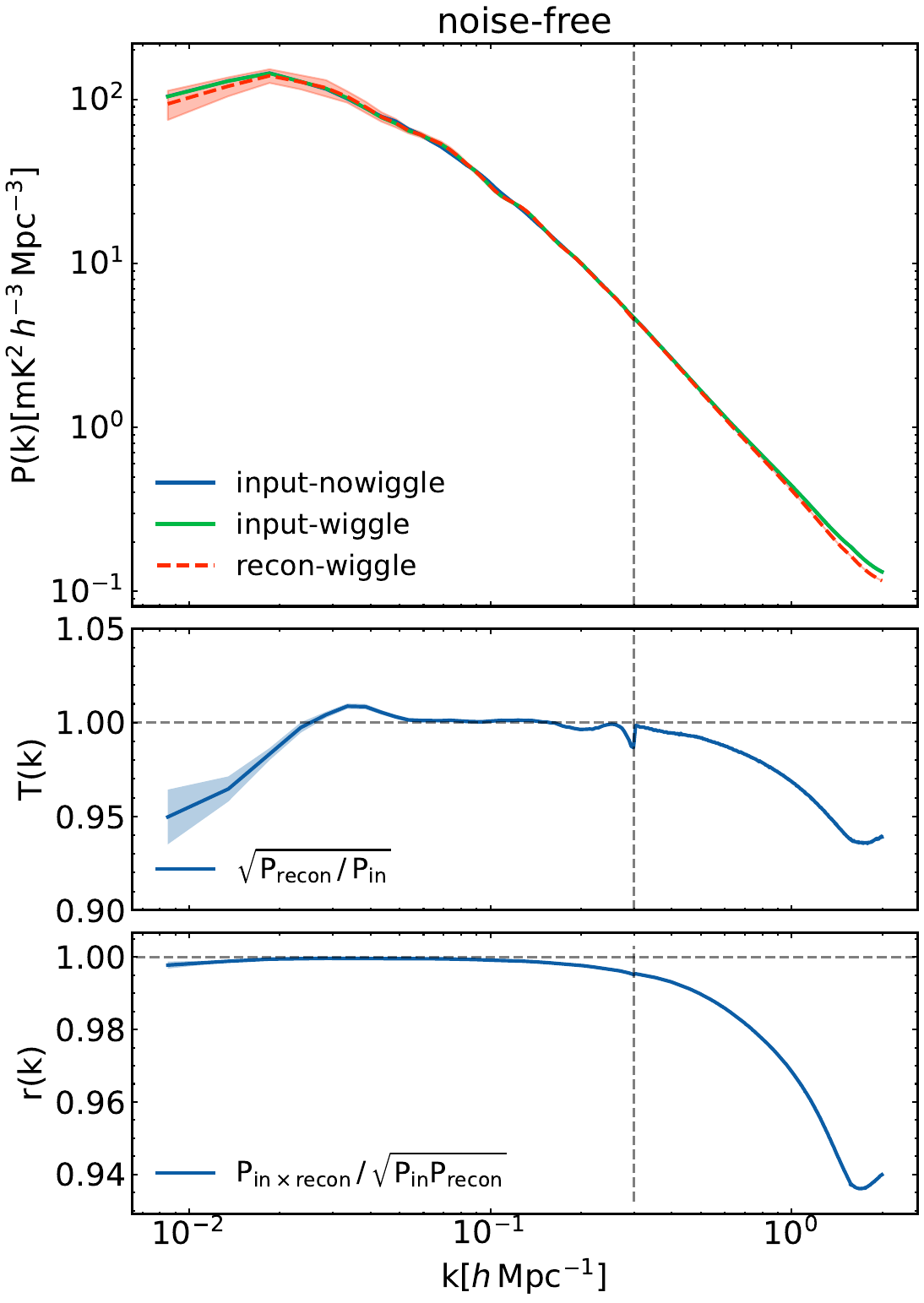}
    \includegraphics[width=0.45\linewidth]{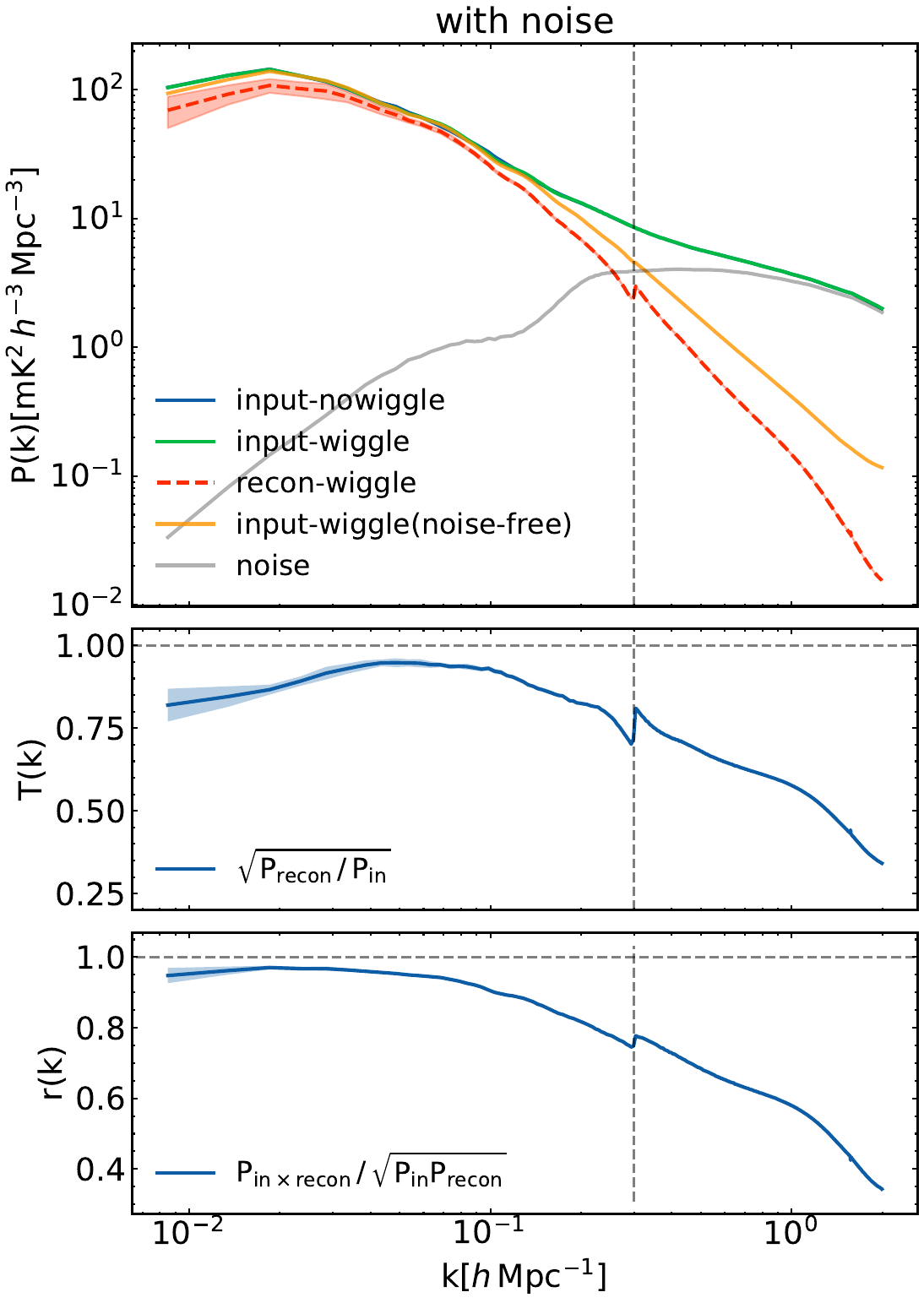}
    \caption{The spherically averaged power spectra (\emph{top}), transfer function (\emph{middle}), and cross-correlation coefficient (\emph{bottom}) for the noise-free (\emph{left}), and the noisy (\emph{right}) $21\,\mathrm{cm}$ brightness temperature fields. \blue{The results shown here averaged over $20$ test realizations, with the shaded regions indicating the $\pm 1 \sigma$ standard deviation. The lines labeled `input-' denote the original fields before modes filtering, while those labeled `recon-' correspond to the fields reconstructed by the trained model from the mode-filtered inputs.} The vertical dashed lines therein indicate the cutoff scale below which modes with $k < 0.3 \,h{\rm Mpc}^{-1}$ are removed from the fields when performing reconstruction using the network. Specifically, in the top right panel, we additionally show the power spectra of the noise-free input field (\emph{orange} line) and noise data (\emph{gray} line).}
    \label{fig:pk-result}
\end{figure}

\begin{figure}
    \centering
    \includegraphics[width=0.4\linewidth]{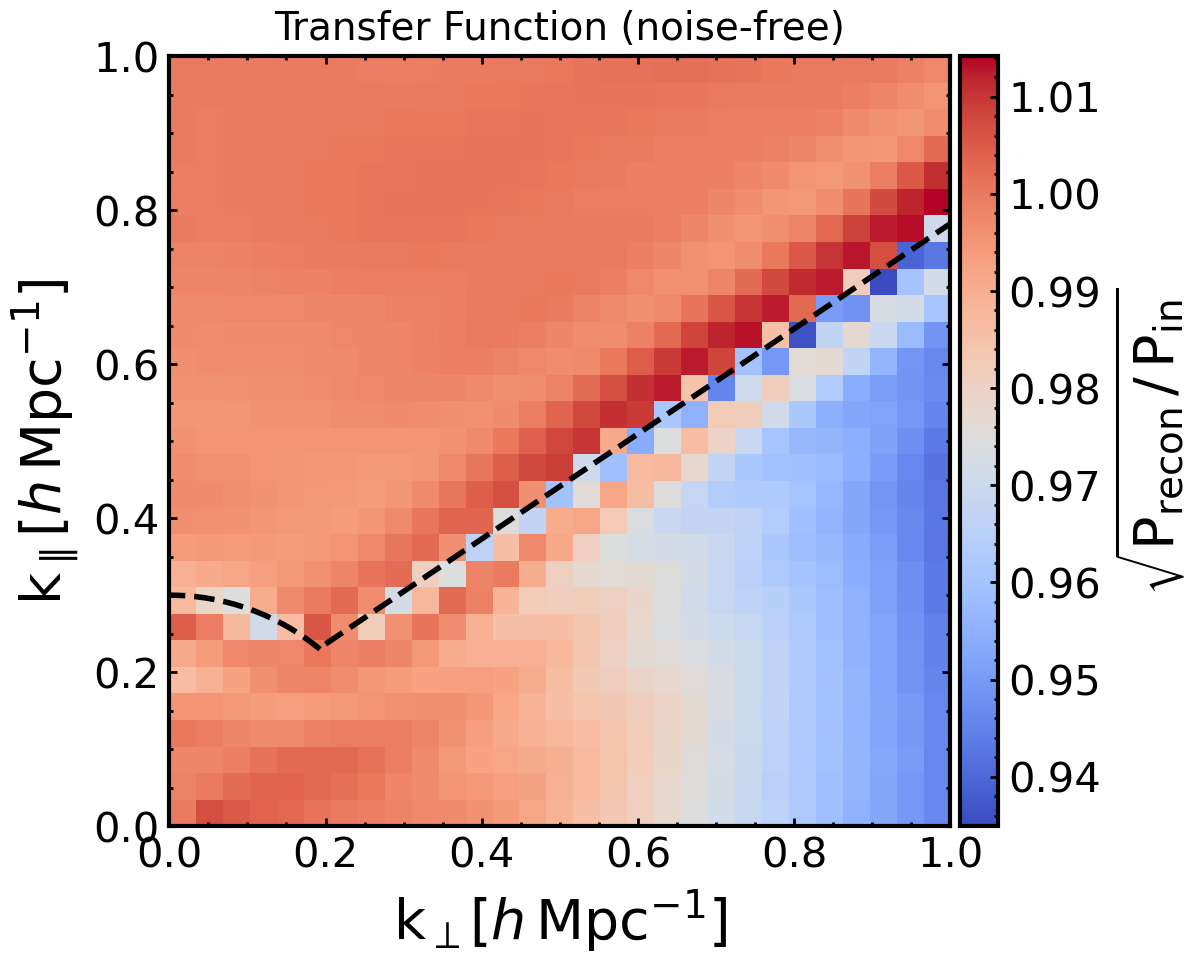}
    \hspace{0.02\linewidth}
    \includegraphics[width=0.4\linewidth]{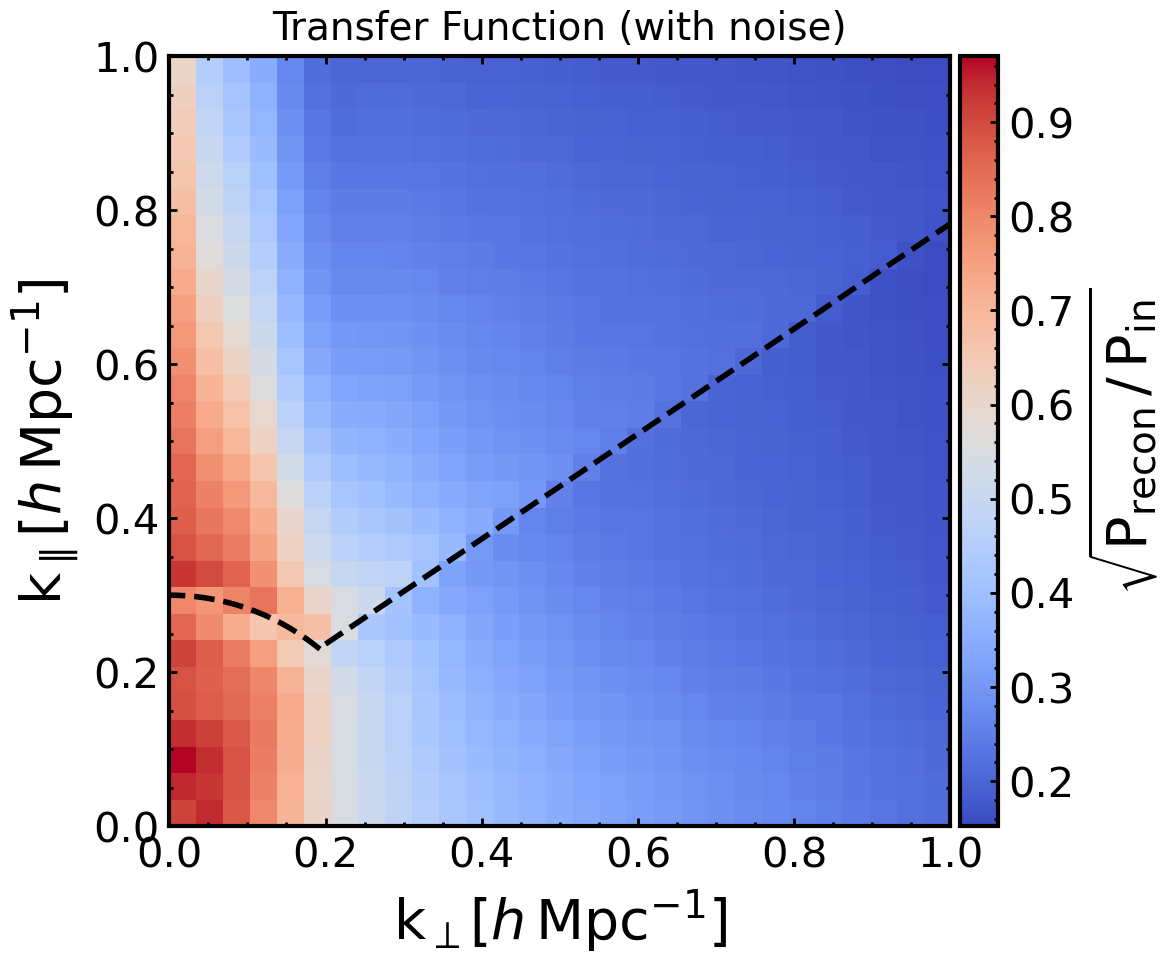} \\
    \includegraphics[width=0.4\linewidth]{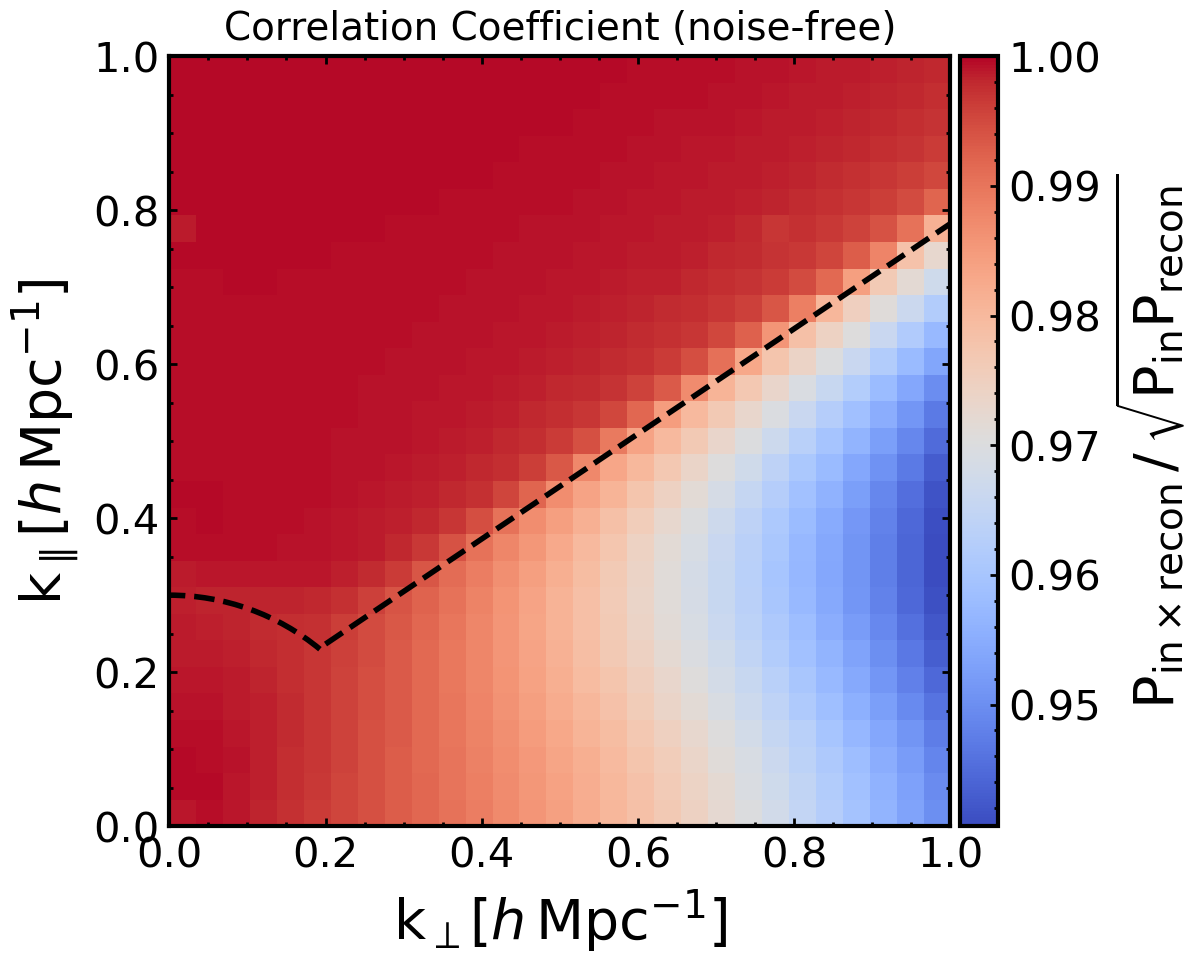}
    \hspace{0.02\linewidth}
    \includegraphics[width=0.4\linewidth]{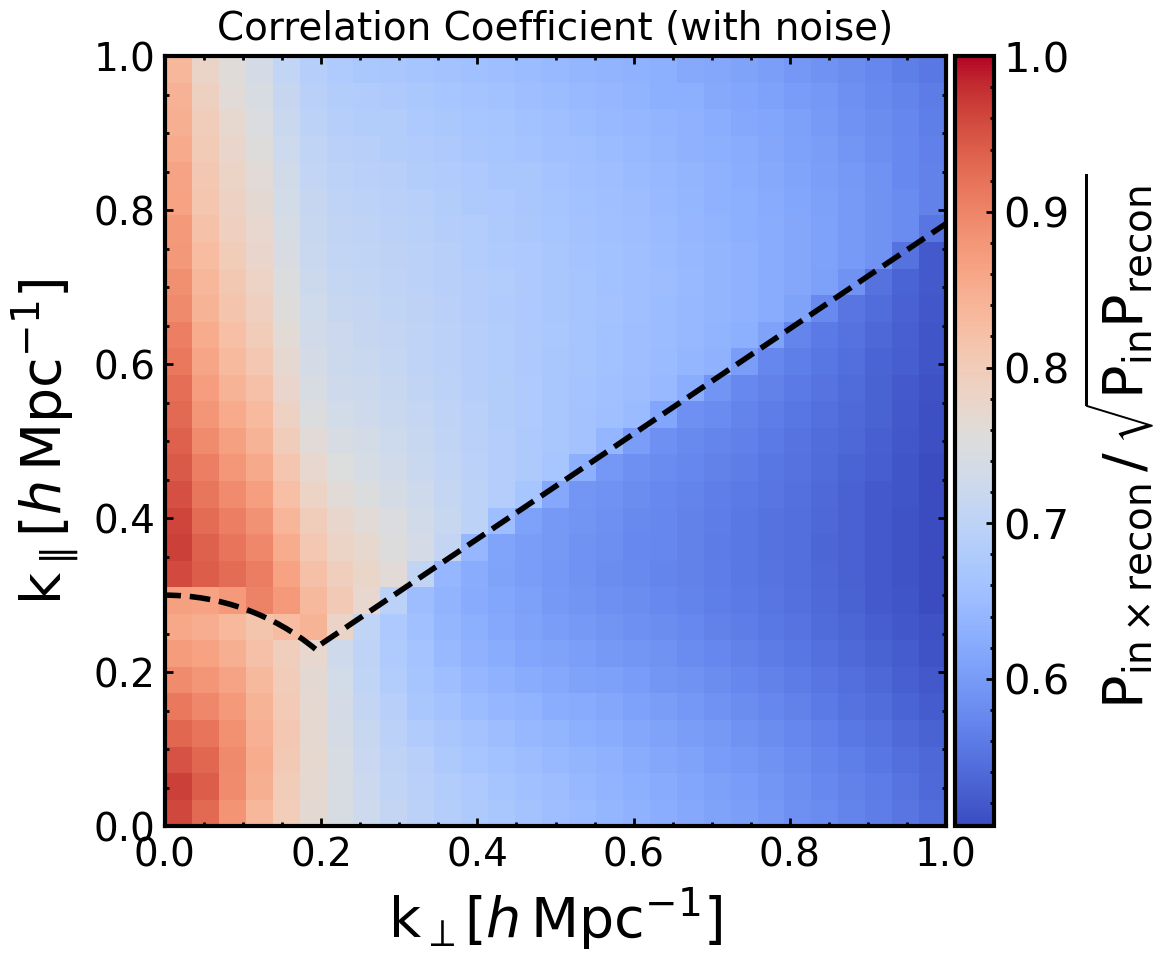}
    \caption{The 2D cylindrical transfer function (\emph{top}) and cross-correlation coefficient (\emph{bottom}) between the input and the reconstructed field for the noise-free (\emph{left}) and noisy data (\emph{right}), \blue{averaged over test datasets}.}
    \label{fig:pk2d-correlation}
\end{figure}

We emphasize here that the network is trained exclusively on the de-wiggled data, and the trained model is subsequently applied to data containing BAO wiggles. Figure~\ref{fig:pk-result} illustrates these metrics evaluated using the spherically averaged one-dimensional power spectrum of the reconstructed field shown in Figure~\ref{fig:nn-recon}.
For the noise-free case, shown in the left panels of Figure~\ref{fig:pk-result}, the power spectrum of the reconstructed field closely follows that of the input field, with noticeable deviations only appearing on the largest and smallest scales. As also indicated by the corresponding transfer function and cross-correlation coefficient, the reconstruction exhibits a reduced fidelity on largest scales $k \lesssim 0.03 \,h{\rm Mpc}^{-1}$ and on scales above the cutoff at $k = 0.3 \,h{\rm Mpc}^{-1}$. 
This behavior reflects the removal of the wedge-shaped region in Fourier space, the fraction of missing modes increases toward higher $k$, which lowers the spherically averaged correlation of the reconstructed field on small scales. The effect is more clearly seen in the two-dimensional transfer function and correlation coefficient shown in Figure~\ref{fig:pk2d-correlation}.
Nevertheless, both the transfer function and the correlation coefficient remain above $0.9$ across all scales, and are close to unity over most of the $k$ range.

For the case with observational noise, shown in the right panels of Figure~\ref{fig:pk-result}, the noise power spectrum exceeds that of the $21\,\mathrm{cm}$ signal at scales of $k \gtrsim 0.3 \,h{\rm Mpc}^{-1}$. As a result, there is an overall suppression of the power spectrum of the reconstructed field relative to the underlying field, with a clear discontinuity at the cutoff scale. In addition, the transfer function and cross-correlation coefficient decline more rapidly toward smaller scales than in the noise-free case.
This behavior is also evident in the reconstructed fields shown in the last column of Figure~\ref{fig:nn-recon}, where the images appear smoother and small-scale structures are suppressed. Since the network is trained to recover a noiseless target field, it naturally suppresses noise-dominated modes, leading to a loss of small-scale information.

\begin{figure}
    \centering
    \includegraphics[width=0.9\linewidth]{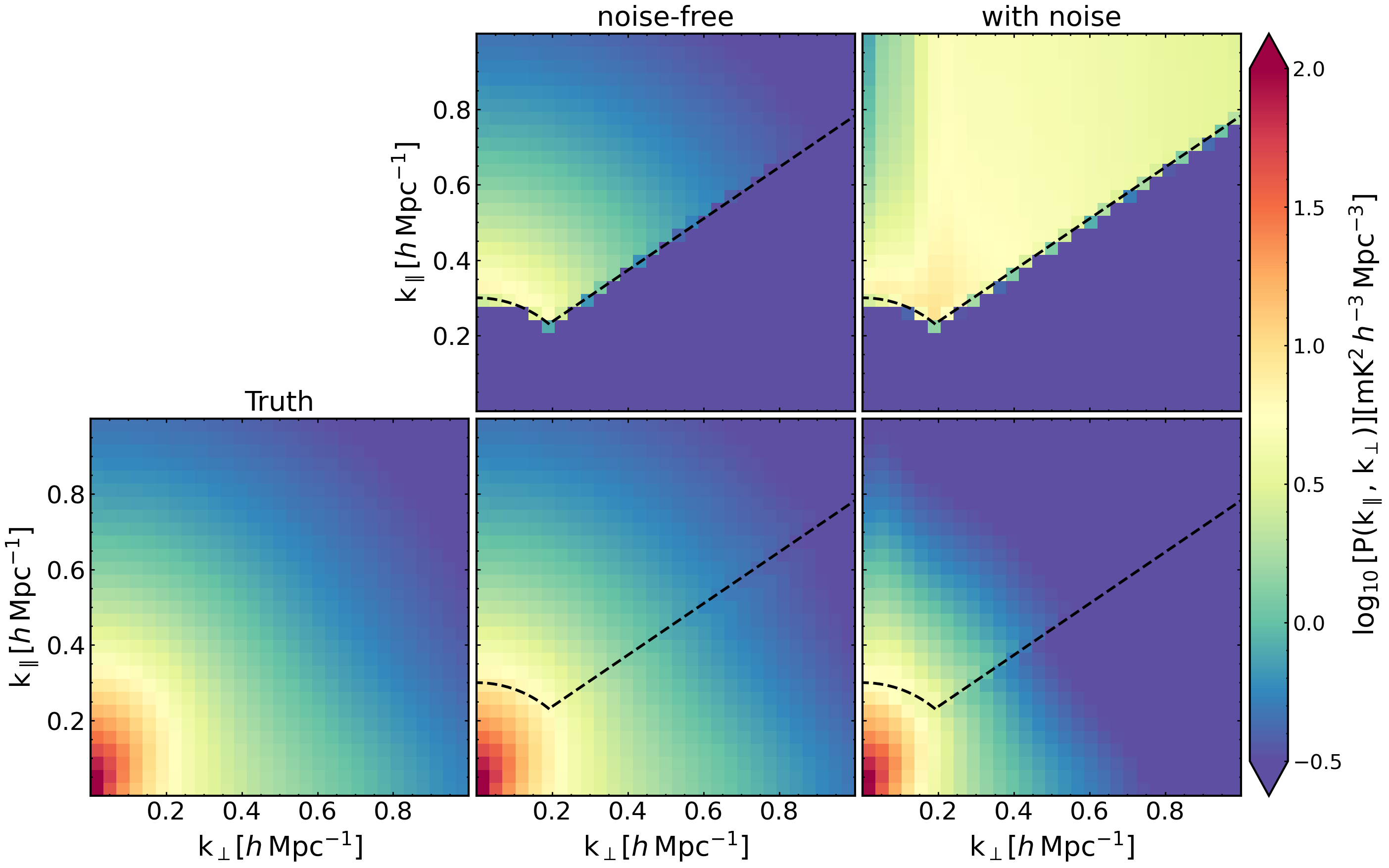}
    \caption{The 2D cylindrical averaged power spectra of the true (\emph{left}), the noise-free (\emph{middle}), and the noisy (\emph{right}) $21\,\mathrm{cm}$ brightness temperature fields, \blue{averaged over test datasets}. \emph{Top}: the power spectrum for the foreground-wedge removed fields. \emph{Bottom}: The corresponding reconstructed fields. The \emph{black} lines therein indicate the boundary of the modes removed ($k < 0.3 \,h{\rm Mpc}^{-1}$ and foreground wedge) in Fourier space.}
    \label{fig:pk2d-result}
\end{figure}

The two-dimensional cylindrical power spectra of the input and reconstructed fields are shown in Figure~\ref{fig:pk2d-result}, providing a clearer view of the reconstruction performance across different $(k_{\perp},\,k_{\parallel})$.
Figure~\ref{fig:pk2d-correlation} presents the corresponding transfer function and cross-correlation coefficient. For the ideal noise-free case, both quantities remain close to unity over most of Fourier space, with noticeable degradation only near the cutoff scale (indicated by the black dashed line) and at relatively large $k_{\perp}$, corresponding to small angular scales. 
This behavior is consistent with the expectation that deeply non-linear small-scale modes are more complex, making the recovery of missing modes increasingly challenging and reducing the effectiveness of the reconstruction at high $k_{\perp}$.
In the presence of observational noise, the reconstruction performance degrades significantly. Nevertheless, modes with $k_{\perp} \lesssim 0.2\,h\mathrm{Mpc}^{-1}$ are still recovered with relatively high fidelity, which is sufficient for extracting the BAO features of the field \blue{as will be demonstrated in Section \ref{sec:result-bao-fitting}}.

%%%%%
\subsection{BAO Signature}
\label{sec:result-bao-fitting}

To evaluate the BAO recovery from the power spectra of the reconstructed $21\,\mathrm{cm}$ fields, we adopt two complementary approaches. 
In the first `baseline' approach, fields with and without BAO wiggles are both passed through the trained network, and the BAO signal is isolated by taking the ratio of their reconstructed power spectra. While this method provides a direct validation of BAO recovery, it relies on access to the de-wiggled fields, which are not available in realistic observations.
Therefore, we also employ a second approach based on template fitting. The fitting method follows the procedure described in \cite{2016MNRAS.460.4210G}, which we briefly summarize here. The linear power spectrum model is decomposed into two components as
\begin{equation}
\label{eq:pk_model}
    P_{\rm model}(k) = O_{\rm damp}(k) P_{\rm model,\, sm}(k).
\end{equation}
Here $O_{\rm damp}(k)$ encodes the BAO oscillating features and is modeled as
\begin{equation*}
    O_{\rm damp}(k) = 1 + [O_{\rm lin}(k / \alpha) - 1] e^{-k^2 \Sigma_{\rm nl}^2 / 2},
\end{equation*}
and the smooth component $P_{\rm model,\, sm}(k)$ is parameterized as
\begin{equation*}
    P_{\rm model,\, sm}(k) = B^2 P_{\rm nw}(k) + \sum^{5}_{i=1}A_i k^{2-i},
\end{equation*}
where $\alpha$, $\Sigma_{\rm nl}$, $B$ and $A_i$ are free parameters to be fitted. $\Sigma_{\rm nl}$ characterizes the non-linear damping of the BAO features, and $O_{\rm lin}$ is the template defined as the ratio between linear power spectrum $P_{\rm lin}(k)$ and its de-wiggled smooth  component $P_{\rm nw}(k)$, i.e. $O_{\rm lin}(k) = P_{\rm lin}(k) / P_{\rm nw}(k)$, which can be obtained using the approach described in Section \ref{sec:nbody-sim}. The parameter $\alpha$ in $O_{\rm lin}(k/\alpha)$ describes the position of BAO peaks.

\begin{figure}
    \centering
    \includegraphics[width=0.4\linewidth]{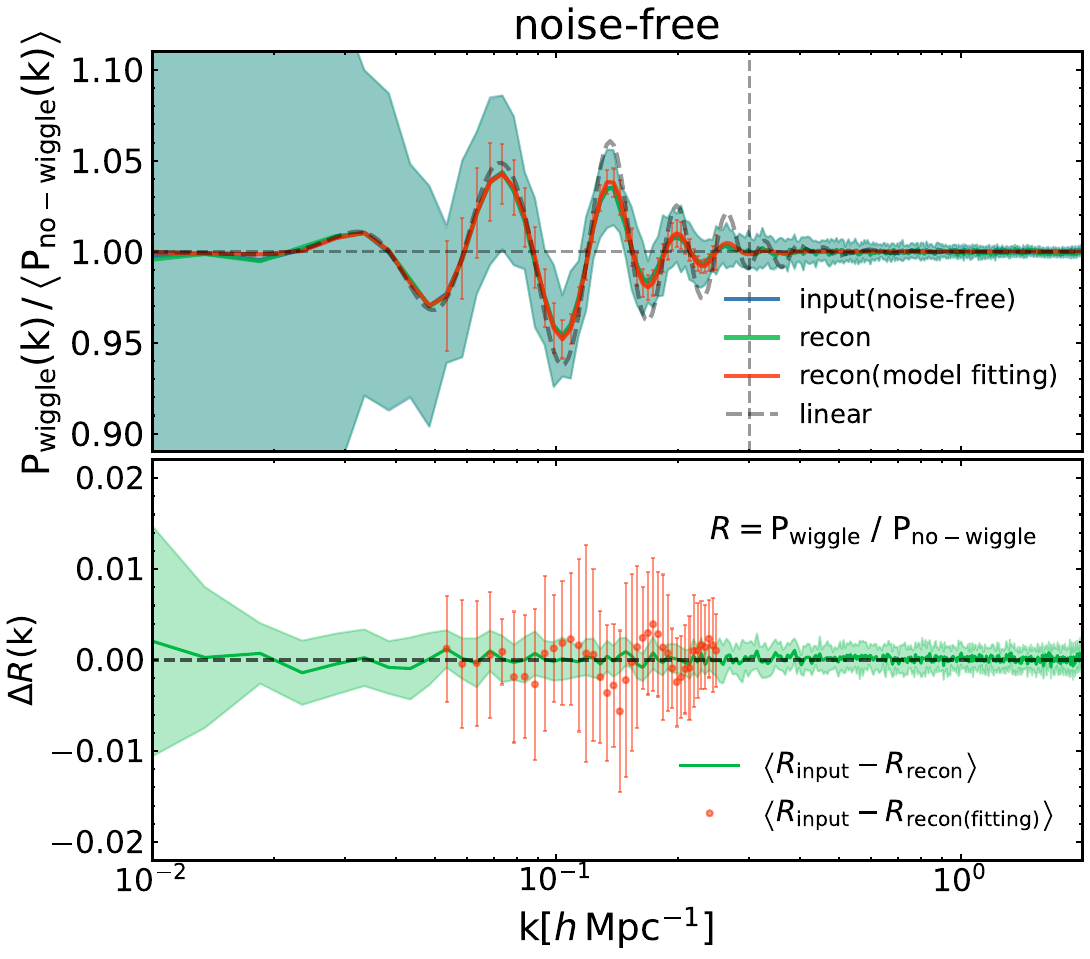}
    \hspace{0.05\linewidth}
    \includegraphics[width=0.4\linewidth]{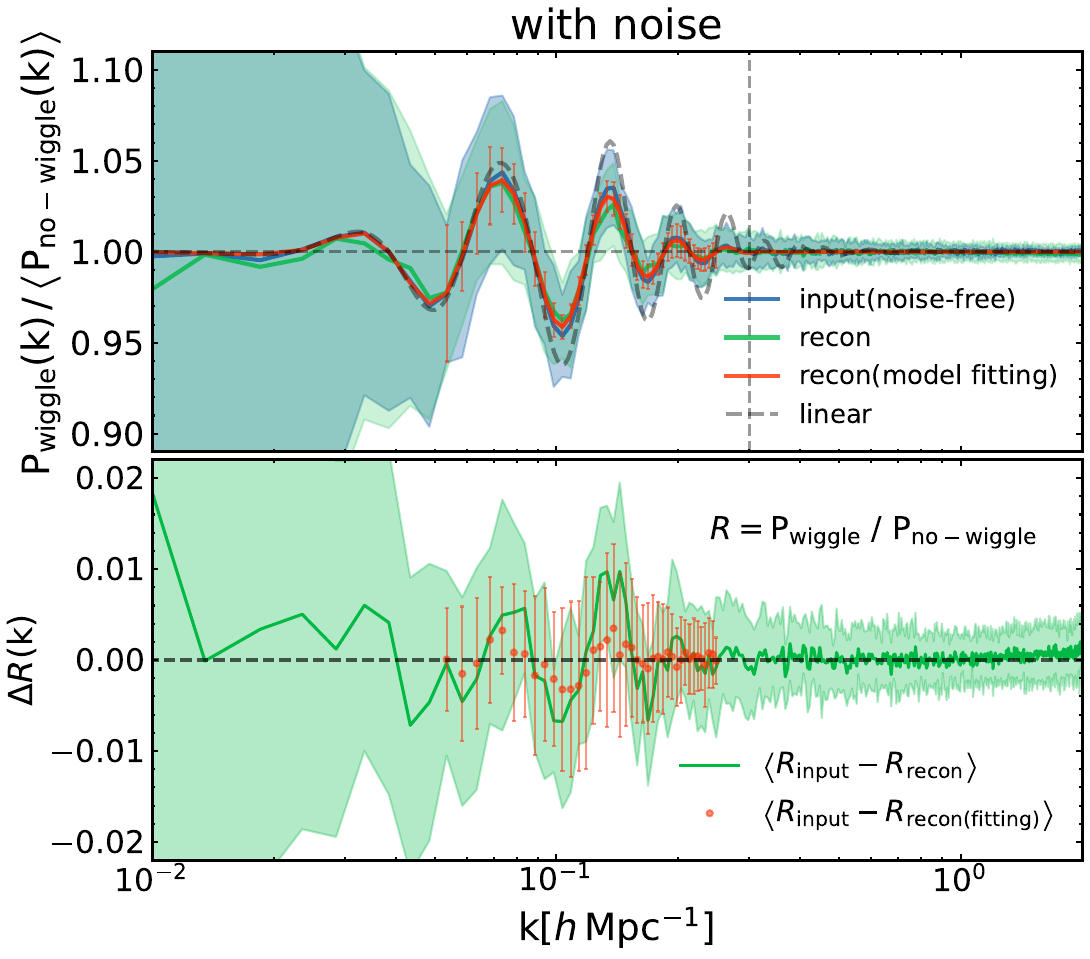}
    \caption{The BAO signature for the noise-free (\emph{left}) and noisy reconstructions (\emph{right}). \blue{\emph{Top}: Ratio of the power spectra between fields with and without BAO wiggles, $P_{\rm wiggle}/\langle P_{\rm no\text{-}wiggle}\rangle$, averaged over $20$ test realizations. Here $\langle P_{\rm no\text{-}wiggle}\rangle$ denotes the mean no-wiggle power spectrum over the test set.} The \emph{blue} curves denote the ratios between input fields with wiggle and no-wiggle before modes removal, \blue{while the \emph{green} curves show the corresponding ratios for the reconstructed fields obtained from the trained models.} The \emph{red} lines indicate the BAO signal extracted using the template-fitting method applied to the reconstructed field with BAO wiggles alone. \blue{The shaded bands and error bars indicate the standard deviation of $P_{\rm wiggle} / \left< P_{\rm no\text{-}wiggle} \right>$ across the $20$ test realizations.} For reference, the BAO wiggle of the linear power spectrum is presented as a \emph{gray} dashed line in each panel. The vertical dashed line indicates the cutoff scale $k = 0.3 \,h{\rm Mpc}^{-1}$ below which modes have been removed from the fields when performing reconstruction.  \blue{\emph{Bottom}: Residuals of the realization-wise BAO ratio, $\Delta R(k)=R_{\rm input}(k)-R_{\rm recon}(k)$, where $R(k)=P_{\rm wiggle}/P_{\rm no\text{-}wiggle}$ is computed from paired wiggle and no-wiggle realizations with the same initial phases. The \emph{green} curves show the residuals for the direct reconstructed ratio, while the \emph{red} points show the residuals obtained from the template-fitting results. The shaded bands and error bars denote the standard deviation across the $20$ test realizations.
    }}
    \label{fig:bao-recon}
\end{figure}

Figure~\ref{fig:bao-recon} illustrates the restored BAO signatures in the reconstructed fields. In both the noise-free and noisy cases, the linear BAO signal is shown as the gray dashed lines for reference. We do not apply any additional BAO reconstruction or linearization to sharpen the signal. Accordingly, both the input and reconstructed BAO features correspond to the non-linearly damped signal.
For the ideal noise-free case shown in the left panel, the BAO signal recovered from the reconstructed fields closely matches that of the original input fields, both in peak positions and amplitudes, demonstrating the efficacy of the reconstruction pipeline. The first three BAO peaks can be clearly identified, independent of whether the BAO signal is extracted using the ratio between reconstructed fields with and without wiggles, or obtained via template-fitting applied to the reconstructed fields with wiggles only. 
At higher orders, the BAO peaks become difficult to identify when using the direct power spectrum ratio, for both the input and reconstructed fields, due to damping from non-linear bulk motions. When using the template-fitting approach, these smaller-scale peaks remain visible by construction, as the oscillatory features are encoded in the fitting template. 
In the presence of observational noise, deviations from the noise-free case become apparent, leading to distortions in the reconstructed BAO features. Nevertheless, the positions of the first two peaks in the reconstructed fields remain clearly identifiable and are consistent with those of the original input fields.

\blue{Figure~\ref{fig:bao-recon} also shows the uncertainty of the recovered BAO signal estimated from the $20$ test realizations. In the upper panels, we show the uncertainty of $P_{\rm wiggle}\,/\,\left< P_{\rm no\text{-}wiggle} \right>$ ({\it blue} and {\it green} shaded region), where we have divided the average `no-wiggle' power spectrum so the scatter arises from cosmic variance and, in the noisy case, thermal noise in the simulated observations. The uncertainties of the input and reconstructed fields are nearly identical in both the noise-free and noisy cases, indicating that the reconstruction does not noticeably increase the realization-to-realization scatter of the BAO ratio. In addition, we also show the scatter of the template-fitted BAO signal ({\it red} error bars). This scatter is smaller than that of the direct power spectrum ratio, mainly because the fitting procedure imposes a smooth oscillatory template and therefore suppresses fluctuations in the measured power spectrum. }

\blue{It does not mean that the recovered modes have exactly the same statistical information as the original full field. In the lower panels, we then provide a consistency check by showing the residual $\Delta R(k) = R_{\rm input}(k)-R_{\rm recon}(k)$, with  $R(k) = P_{\rm wiggle}\,/\, P_{\rm no\text{-}wiggle}$ computed for paired wiggle and no-wiggle realizations that share the same initial phases. This pairing largely cancels cosmic variance and makes the residual sensitive to reconstruction-induced differences. For the direct ratio in the noise-free case, the residual scatter is small ({\it green} shaded region),  at the level of $\sim 0.2\%$ around $k\simeq 0.1\,h\mathrm{Mpc^{-1}}$. This indicates that the network does introduce a small realization-dependent variation when reconstructing fields generated from different input power spectra.
For the template-fitting result ({\it red} error bars), the scatter in the residual is larger because the BAO signal is extracted indirectly through the fitted template component $O_{\rm damp}(k)$ in Equation~\eqref{eq:pk_model} using only the wiggle realization without reference to the corresponding no-wiggle field. Small differences between the input and reconstructed power spectra can shift the best-fit parameters, and these shifts propagate into $O_{\rm damp}(k)$. In addition, the broadband and oscillatory components are not completely independent in the fit. Therefore, although the fitted BAO curves are smoother in the upper panels, their realization-to-realization residuals can be larger than those obtained from the direct paired ratio.
For the noisy case, since the same thermal noise realization is added to each paired wiggle and no-wiggle fields, the increased scatter in the direct-ratio residual mainly arises from the way noise propagates through the non-linear reconstruction process. 
}

\blue{
In summary, for the noise level considered in this work, the variance is dominated by cosmic variance and thermal noise. Around $k\simeq 0.1\,h\mathrm{Mpc}^{-1}$, this contribution is larger than the reconstruction induced scatter by a factor of $\sim 4$. The additional scatter introduced by the reconstruction is therefore subdominant for the BAO statistic considered here. We note, however, that this conclusion is based on only $20$ simulated realizations. A more rigorous assessment using a larger ensemble of simulations and a more detailed noise treatment is left for future work.
}

%%%%%
\subsection{Robustness of the Network}
\label{sec:robustness-nn}

\begin{figure}
    \centering
    \includegraphics[width=0.4\linewidth]{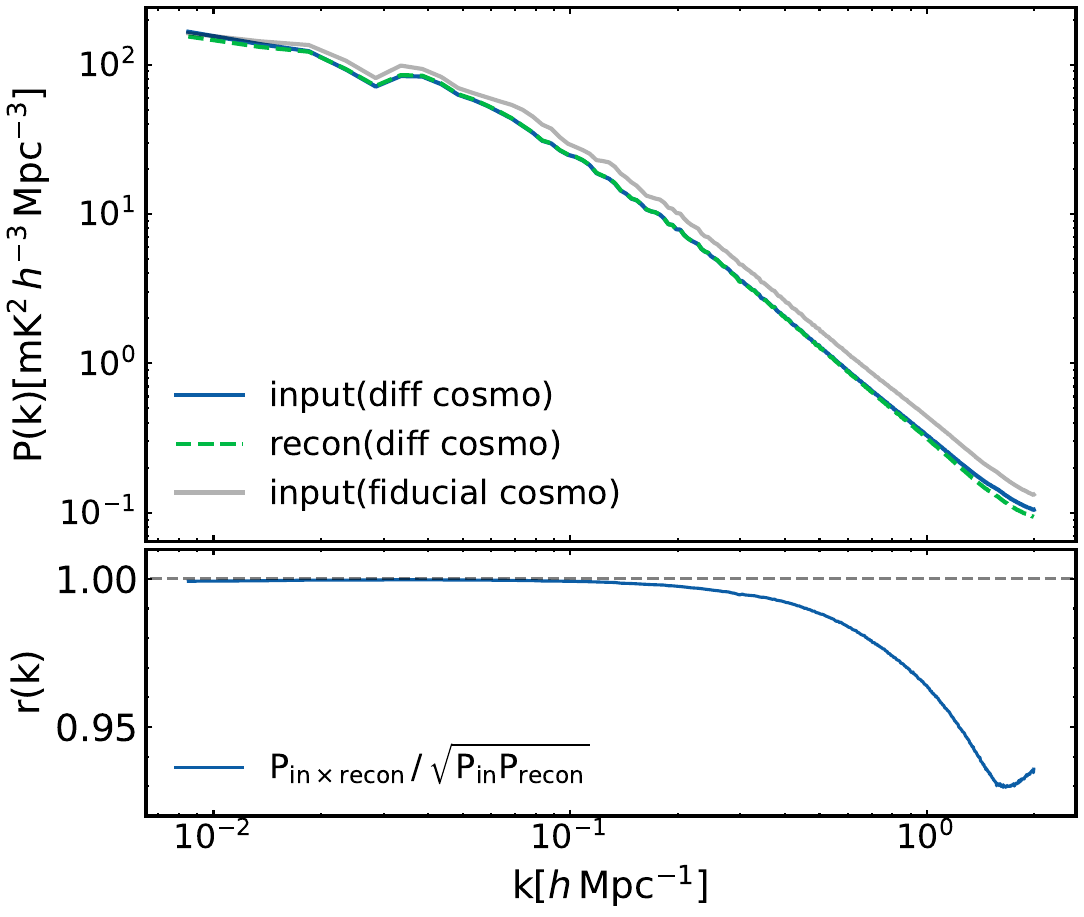}
    \hspace{0.05\linewidth}
    \includegraphics[width=0.4\linewidth]{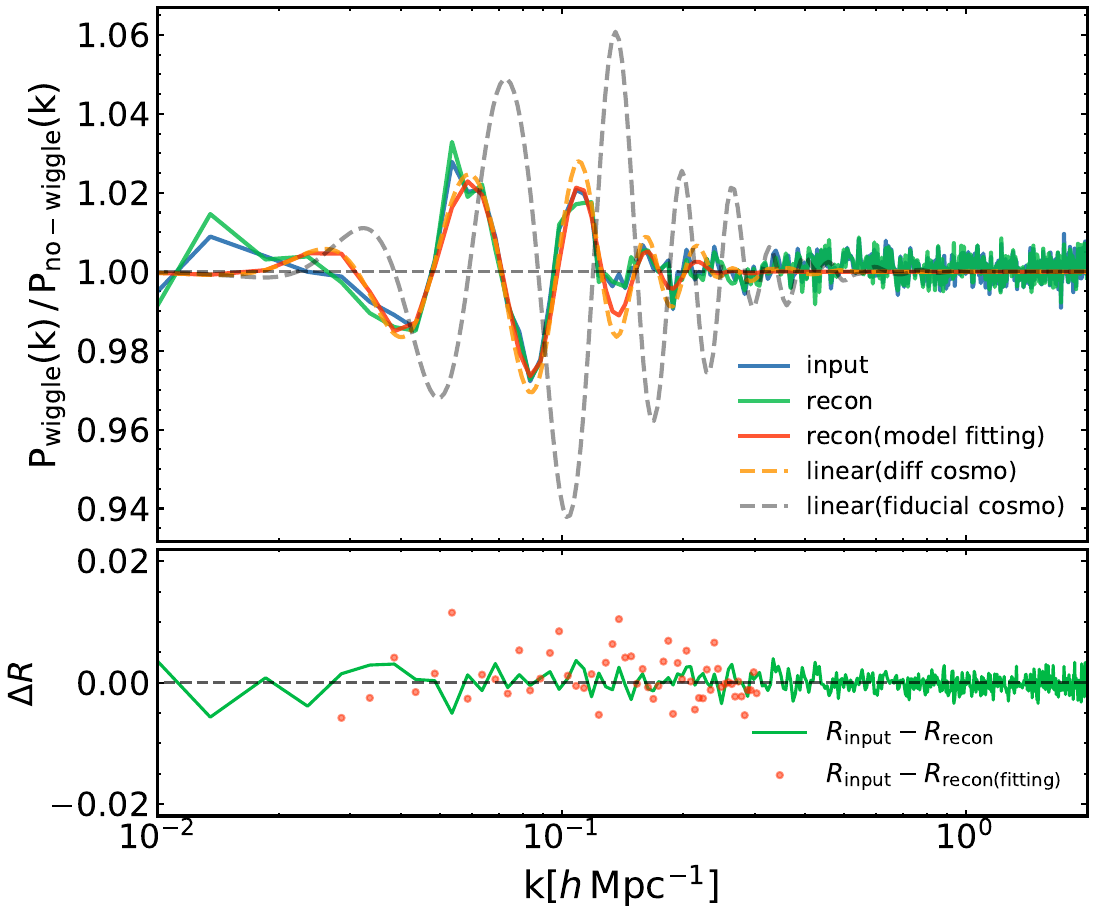}
    \caption{The 1D power spectrum (\emph{left}) and the BAO features (\emph{right}) of the reconstructed $21\,\mathrm{cm}$ field evolved under a different cosmology. \blue{The \emph{bottom} panel of the right figure shows the residual in BAO features between the input and the reconstructions.} In this setup, the cosmological parameters differ only in the $N$-body simulations, while they are kept identical to the fiducial one in the subsequent computation of the $21\,\mathrm{cm}$ fields described in Section \ref{sec:HI-Tb}. The \emph{gray} line in the left panel represents the 1D power spectrum of the fiducial field sharing the same initial condition seed.}
    \label{fig:bao-fitting-diff-cosmo}
\end{figure}

\begin{figure}
    \centering
    \includegraphics[width=0.4\linewidth]{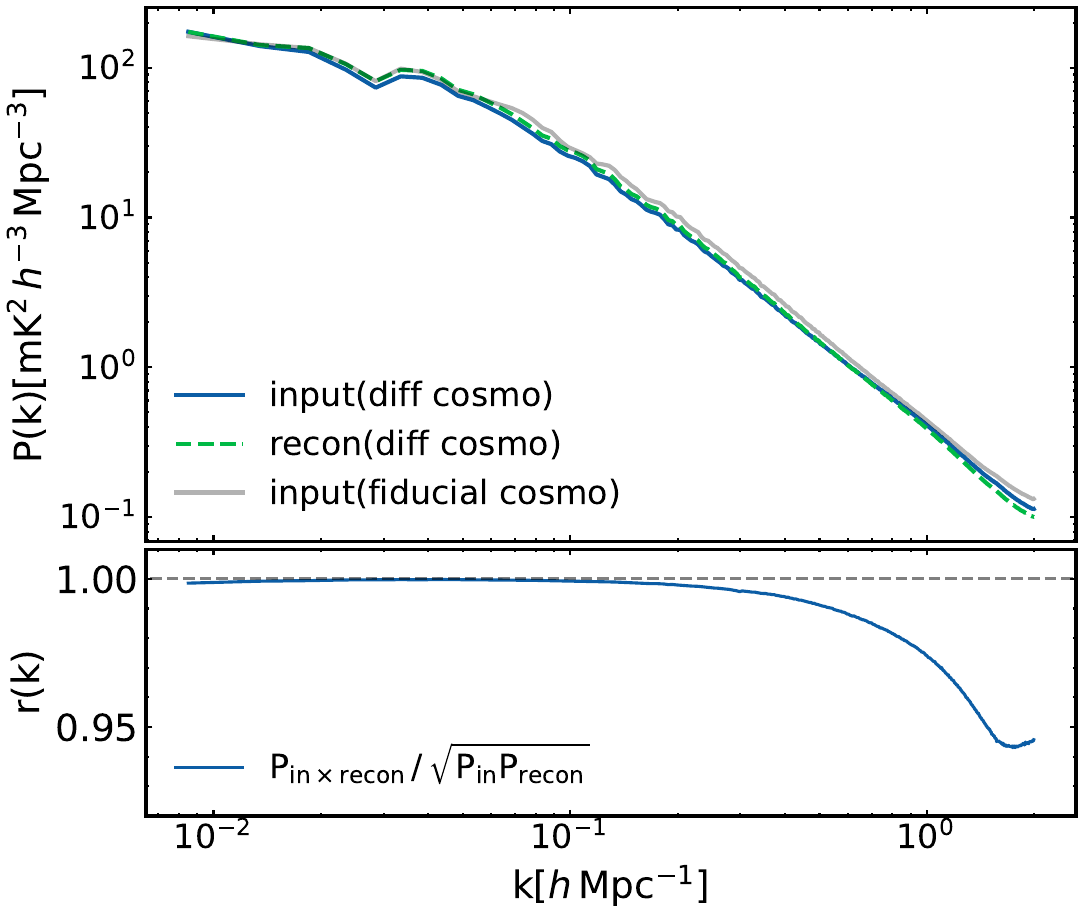}
    \hspace{0.05\linewidth}
    \includegraphics[width=0.4\linewidth]{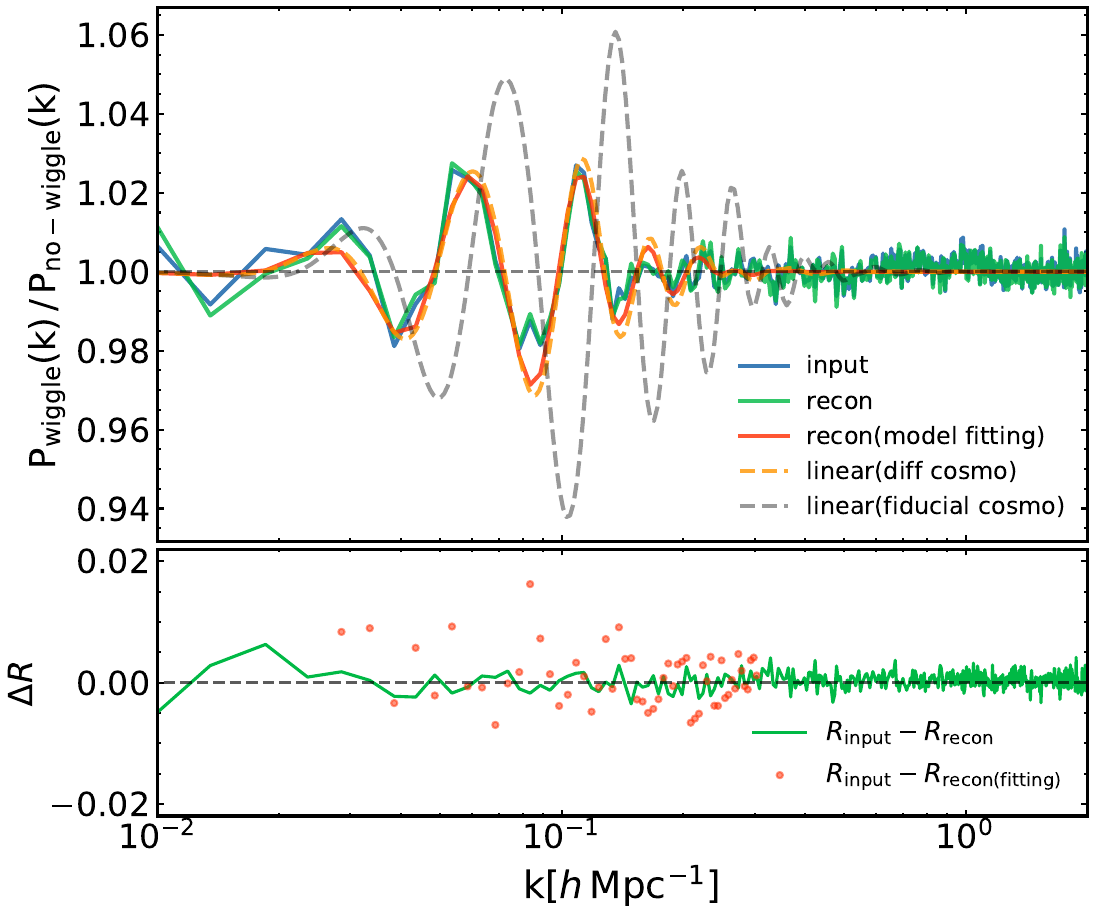}
    \caption{Same as Figure~\ref{fig:bao-fitting-diff-cosmo}. In this case, the set of cosmological parameters in the production of the $21\,\mathrm{cm}$ temperature fields is different from the fiducial one as well.}
    \label{fig:bao-fitting-diff-cosmo-2}
\end{figure}

\begin{figure}
    \centering
    \includegraphics[width=0.4\linewidth]{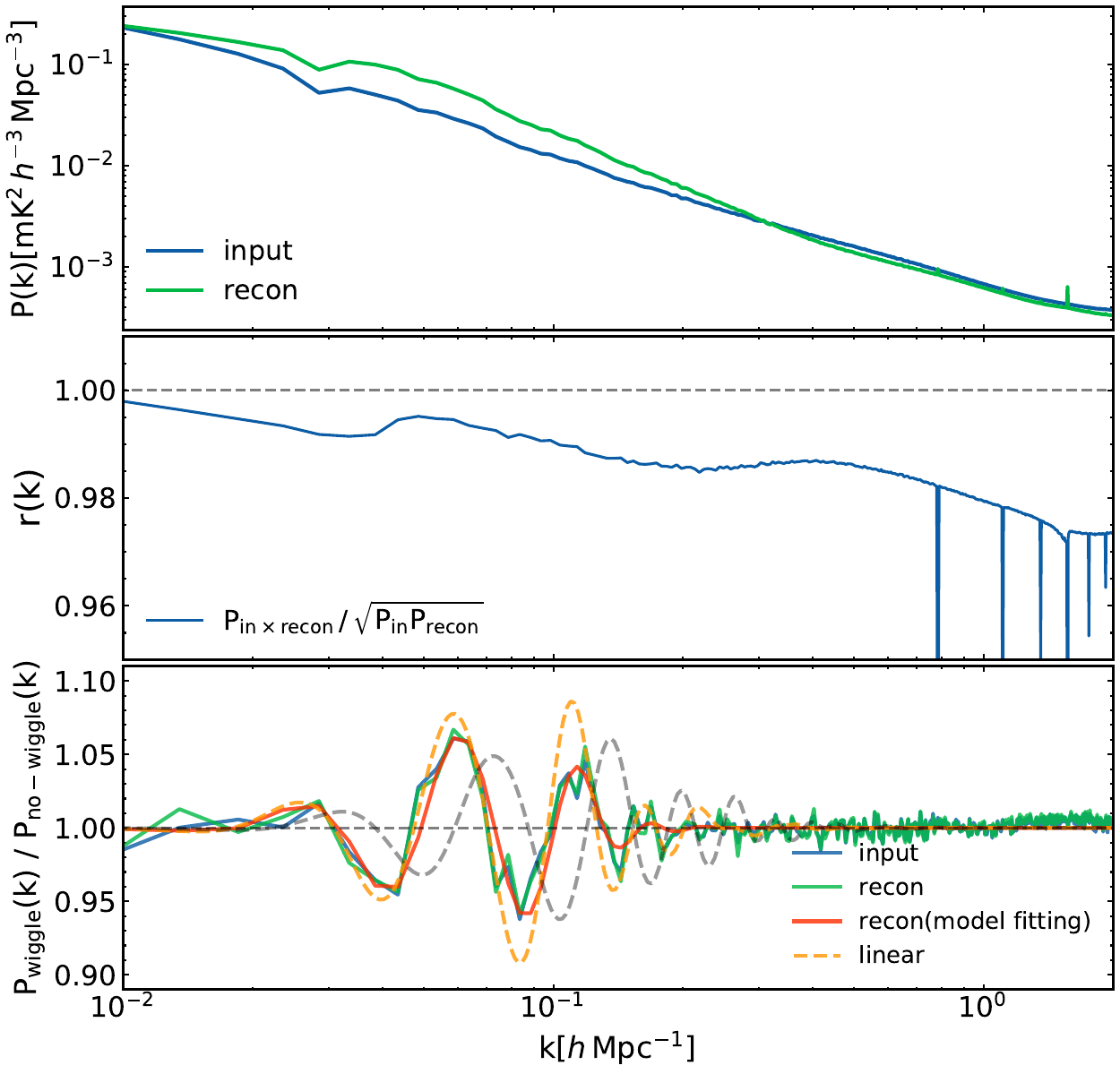}
    \hspace{0.05\linewidth}
    \includegraphics[width=0.4\linewidth]{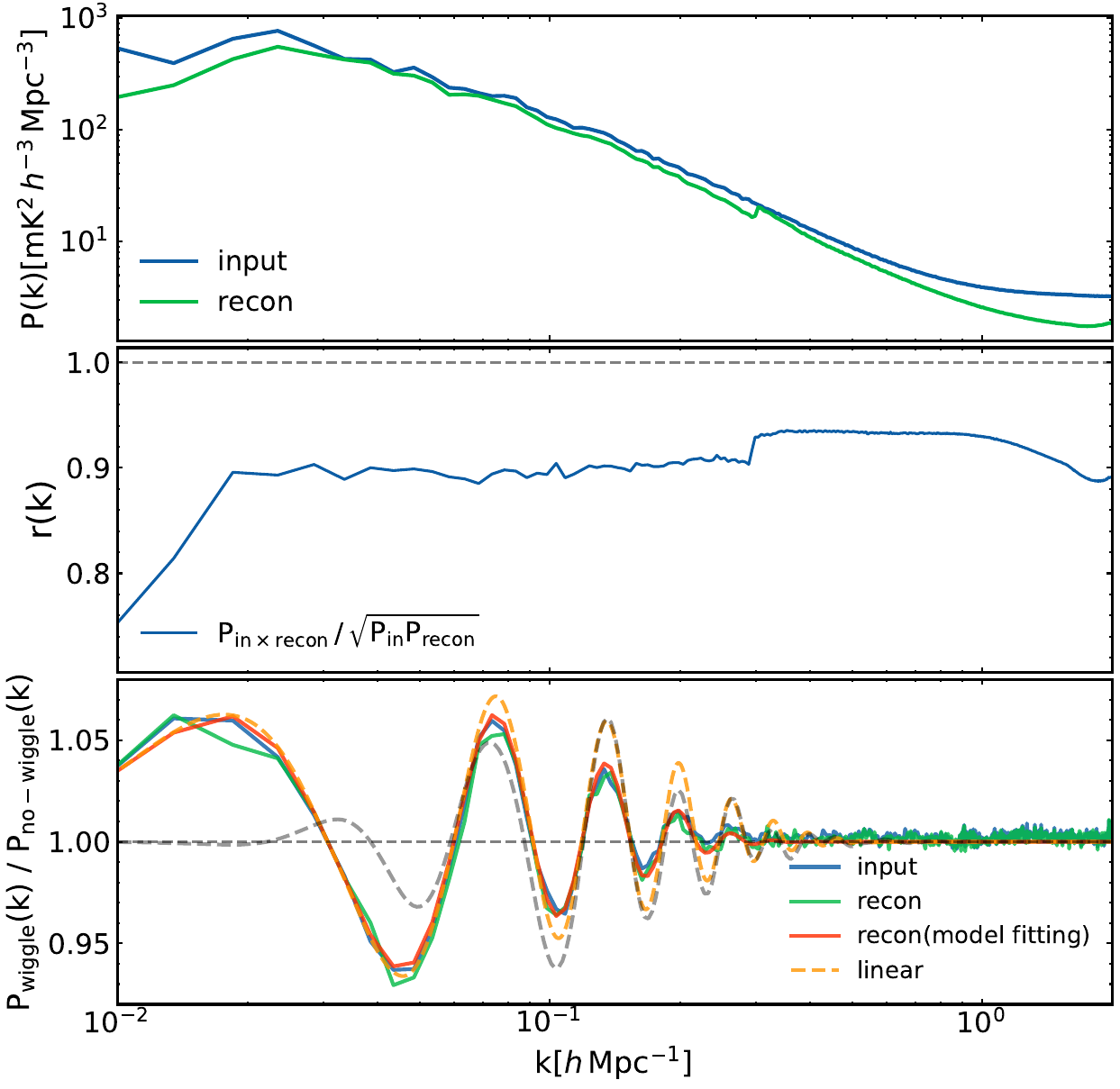}
    \caption{\blue{\emph{Left}: Same as Figure~\ref{fig:bao-fitting-diff-cosmo-2}, but for another cosmology with $\Omega_{\rm m} = 0.12$, $h = 0.55$, and $\sigma_8 = 0.35$. \emph{Right}: Results for data generated from GADGET simulations. The detailed description of the data can be found in  \cite{2025ApJ...983..166L}. The \emph{gray} lines in the bottom panels indicate the fiducial linear BAO features used to generate the training datasets.}}
    \label{fig:bao-fitting-diff-cosmo-4}
\end{figure}

The training data are generated using a single set of cosmological parameters. To assess the robustness of the trained model to variations in the underlying cosmology, and to test whether it captures the relevant mode coupling rather than features specific to the training set, we generate an additional pair of $N$-body simulations with a different cosmological model with $\Omega_{\rm m} = 0.22$, $h = 0.72$, and $\sigma_8 = 0.834$. For the subsequent mapping from dark matter to $21\,\mathrm{cm}$ brightness temperature fields, we consider two modeling choices. In one case, the $21\,\mathrm{cm}$ field is generated using the fiducial Planck 2018 cosmological parameters, while in the other case it is constructed using parameters consistent with those adopted in the $N$-body simulations. The corresponding reconstruction results are shown in Figures~\ref{fig:bao-fitting-diff-cosmo} and \ref{fig:bao-fitting-diff-cosmo-2}, respectively. \blue{We further tested an additional set of cosmological parameters with $\Omega_{\rm m} = 0.12$, $h = 0.55$, and $\sigma_8 = 0.35$, and the results are shown in the left panel of Figure~\ref{fig:bao-fitting-diff-cosmo-4}. Meanwhile, we apply the model trained on datasets generated from COLA simulations to data produced from GADGET simulations, with the corresponding results presented in the right panel of Figure~\ref{fig:bao-fitting-diff-cosmo-4}.}

We focus exclusively on the noise-free case here. For scenarios including observational noise, we find that when the noise level is consistent with that adopted in previous analyzes, the conclusions remain similar to those obtained in fiducial cases discussed above.
For the case with an alternative $N$-body cosmology combined with the fiducial $21\,\mathrm{cm}$ model, shown in Figure~\ref{fig:bao-fitting-diff-cosmo}, the left panel demonstrates that although the underlying power spectrum differs from that of the fiducial cosmology, the reconstructed field still recovers both amplitude and phase information with high fidelity. This is reflected by the close agreement between the reconstructed and input power spectrum, as well as the high cross-correlation coefficients. Regarding the extraction of BAO wiggle features, as illustrated in the right panel, we find that even though the BAO features in the training set differ from those in this scenario, the trained model is able to recover the underlying BAO feature encoded in the field from those non-linear modes.

The reconstruction results for simulations using an alternative cosmological model for both the $N$-body and subsequent $21\,\mathrm{cm}$ modeling are shown in Figure~\ref{fig:bao-fitting-diff-cosmo-2} \blue{and Figure~\ref{fig:bao-fitting-diff-cosmo-4}}. In \blue{these} cases, the reconstructed power spectra show a noticeable amplitude mismatch relative to the input fields, rather than the near-perfect agreement seen in the previous case.  Despite this amplitude discrepancy, the cross-correlation coefficients remain comparable in both magnitude and scale dependence, indicating that the phase information is still well recovered. The BAO extraction shown in the \blue{figures} further demonstrates that the BAO signal can be successfully recovered in \blue{these} scenarios as well. 
These results indicate that the trained network exhibits a degree of robustness to variations in cosmological parameters. Even when the data differ from those represented in the training set, the model is still able to produce reasonable reconstructions. This \blue{may} suggest that the network \blue{captures aspects of} the mode coupling structure in the data, \blue{which could help} it infer missing modes from the information encoded in the remaining ones.

%%% Conclusions
\section{Conclusions}
\label{sec:conclution}

\blue{In the interferometric $21\,\mathrm{cm}$ intensity mapping, the presence of the foreground wedge restricts foreground avoidance strategies to relatively foreground-clean small-scale modes, thereby removing a significant fraction of large-scale information. However, non-linear evolution induces coupling between Fourier modes. As a result, the information contained in a subset of modes can, in principle, be used to recover the missing modes.}
In this work, focusing on post-EoR observations, we train neural networks to reconstruct fields that retain only modes outside the foreground wedge in the non-linear regime, with the goal of recovering the missing modes at the field level. Our results demonstrate that the model is indeed capable of recovering the lost modes, and especially the BAO signature, using information from the available subset of modes alone.
In particular, to verify that the reconstruction is driven by mode coupling information rather than by memorization of specific patterns in the training set, we construct the training set exclusively from data that do not contain BAO features and subsequently apply the trained model to data exhibiting wiggle features. The trained networks are nevertheless able to recover these features that are entirely absent from the training data.
To further assess the robustness of the model under different cosmological scenarios, we perform additional tests using data generated with cosmological parameters different from those adopted in the training set. These tests indicate that the phase information of the reconstructed fields, quantified by the correlation coefficient, can still be recovered with high fidelity, while discrepancies in the reconstructed amplitude persist. A detailed investigation of the origin of this amplitude mismatch is left to future work.

The treatment of the training data in this work remains relatively simplified. In future studies, we might employ higher-resolution hydrodynamical simulations to generate more accurate neutral hydrogen intensity fields for training. In addition, more complex systematic effects, such as leakage beyond the theoretical foreground wedge and instrumental calibration errors, should be incorporated and carefully examined. It will also be necessary to combine observation data in order to further validate and assess strategies for network training, ensuring that they are robust and reliable when applied to real observations.
Nevertheless, as a proof of concept, this study demonstrates the feasibility of using interferometric $21\,\mathrm{cm}$ intensity mapping to recover large-scale structure, particularly the BAO signal, by exploiting non-linear information in the fields.

% \appendix
% \section{Some title}
% Please always give a title also for appendices.

\acknowledgments
We thank the anonymous reviewer for the constructive comments and suggestions.
This work is supported by the National SKA Program of China (Grants Nos. 2022SKA0110200, 2022SKA0110202), the National Science Foundation of China (Grants Nos. 12473006), the China Manned Space Project with No. CMS-CSST-2021 (B01, A02, A03).

% Bibliography

%% [A] Recommended: using JHEP.bst file
\bibliographystyle{JHEP}
\bibliography{reference.bib}

%% or
%% [B] Manual formatting (see below)
%% (i) We suggest to always provide author, title and journal data or doi:
%% in short all the informations that clearly identify a document.
%% (ii) please avoid comments such as "For a review'', "For some examples",
%% "and references therein" or move them in the text. In general, please leave only references in the bibliography and move all
%% accessory text in footnotes.
%% (iii) Also, please have only one work for each \bibitem.

\end{document}